\documentclass[journal=jctcce,manuscript=article]{achemso}
\usepackage[T1]{fontenc} 
\usepackage{amsmath}
\usepackage{amssymb}
\usepackage{bm} 
\usepackage{placeins} 

\newcommand{\sumover}[2]{\ensuremath{\underset{#1}{\overset{#2}{\sum}}}}
\newcommand{\quotes}[1]{``#1''}
\newcommand{\latin}[1]{\textit{#1}}
\author{Nina Glaser}
\author{Alberto Baiardi}
\author{Markus Reiher}
\email{mreiher@ethz.ch}
\affiliation[ETH Zurich]{ETH Zurich, Department of Chemistry and Applied Biosciences, Vladimir--Prelog-Weg 2, 8093 Zurich, Switzerland}

\date{October 16, 2023}

\title[n-mode vDMRG]{Flexible DMRG-Based Framework for Anharmonic Vibrational Calculations}

\abbreviations{DMRG,VSCF,HDMR,vDMRG,SRCAS,VCI}
\keywords{DMRG,$n$-mode PES,VSCF,anharmonic vibrational spectroscopy,vDMRG,SRCAS} 

\SectionNumbersOn

\begin{document}



\begin{abstract}
 
\noindent We present a novel formulation of the vibrational density matrix renormalization group (vDMRG) algorithm tailored to strongly anharmonic molecules described by general high-dimensional model representations of potential energy surfaces.
For this purpose, we extend the vDMRG framework to support vibrational Hamiltonians expressed in the so-called $n$-mode second-quantization formalism.
The resulting $n$-mode vDMRG method offers full flexibility with respect to both the functional form of the PES and the choice of the single-particle basis set.
We leverage this framework to apply, for the first time, vDMRG based on an anharmonic modal basis set optimized with the vibrational self-consistent field algorithm on an on-the-fly constructed PES.
We also extend the $n$-mode vDMRG framework to include excited-state targeting algorithms in order to efficiently calculate anharmonic transition frequencies.
We demonstrate the capabilities of our novel $n$-mode vDMRG framework for methyloxirane, a challenging molecule with 24 coupled vibrational modes.
\end{abstract}

\clearpage
\section{Introduction}

Vibrational spectroscopy is a powerful analytical tool that is routinely applied in various fields of chemistry to elucidate compound structures and to gain insight into molecular properties.\cite{long1977_ramanBook,barron_2004,quack2011handbook,larkin17_spectroscopyBook}
To that end, there is a high demand for accurate predictions of vibrational spectra based on \latin{ab initio} calculations.
However, the highly accurate calculation of vibrational spectra for systems with more than a dozen degrees of freedom remains one of the major challenges in molecular spectroscopy.\cite{puzzarini19_compSpectReview,bowman22_book}
Within the Born--Oppenheimer approximation,\cite{born27} two key steps are required to obtain reliable vibrational spectra:
First, an accurate anharmonic potential energy surface (PES) must be constructed.
Second, the resulting anharmonic many-body vibrational Schr\"{o}dinger equation must be solved with an adequate vibrational structure method.

In the past two decades, a wide variety of techniques have been developed to achieve the first task, \latin{i.e.}, to parametrize the PES for molecules with several dozen vibrational degrees of freedom. This includes approaches based on permutationally invariant polynomials,\cite{Bowman2011_PIP-Review} Gaussian processes,\cite{Csanyi2015_PES-GaussianApproximation,Behler2016_MLPES-Review,Kolb2017_ReactivePES-ML} and neural-networks,\cite{Manzhos2015_NN-PES,Guo2016_PIP-NN,Brorsen2019_NeuralNetwork}.
However, the computational cost associated with the exact solution of the resulting vibrational Schr\"{o}dinger equation scales exponentially with system size, prohibiting straightforward calculations for molecules with more than a dozen degrees of freedom.
The design of novel anharmonic methods to tame this unfavorable scaling remains an active field of research.\cite{carrincton17_compSpectPerspective,madsen21_vcc,tran2023_vhbci}
Among the vibrational structure methods developed, techniques leveraging so-called tensor network-based algorithms are particularly promising.\cite{Manthe2008_MLMCTDH-Original,Leclerc2014_TensorDecomposition,godtliebsen13,Oseledts2016_VDMRG,larsson19}
The most commonly applied tensor-based algorithm in electronic structure theory has been the density matrix renormalization group (DMRG),\cite{white92,White1993_DMRGBasis} which allows for an efficient deterministic variational optimization of wave functions represented as matrix product states (MPSs).\cite{McCulloch2007_FromMPStoDMRG,chan09_dmrgRev,Zgid2009_Review,Marti2010_Review-DMRG,Chan2011_Review,Schollwoeck2011_Review-DMRG,Wouters2013_Review,Kurashige2014_Review,Olivares2015_DMRGInPractice,Szalay2015_Review,Yanai2015,Knecht2016_Chimia,Baiardi2020_Review,ma22_dmrg-book}
In previous work, we introduced the vibrational DMRG (vDMRG) algorithm,\cite{Baiardi2017_VDMRG,Baiardi2019_HighEnergy-vDMRG} which enables large-scale vibrational structure calculations of anharmonic systems described by Taylor-expanded PESs.

Taylor series expansions are the prevalent functional format to efficiently approximate many-body potential energy surfaces since they have the innate benefit to be automatically encoded in a sum-over-products form, which is a convenient parametrization for various vibrational structure approaches.
This is also the case for vDMRG, as the Taylor expansion naturally provides a second-quantization framework in which to express the vibrational Hamiltonian in matrix product operator (MPO) form.\cite{glaser22}
The resulting so-called canonical quantization spans the Hilbert space generated by the harmonic oscillator eigenfunction basis.

While vDMRG in canonical quantization is a powerful approach for studying weakly anharmonic molecules, it suffers from certain fundamental limitations.
As the Taylor expansion leads to a second-quantized form relying on the harmonic oscillator eigenfunction basis, encoding strongly anharmonic wave functions accurately in such a basis requires a high number of harmonic basis functions.
Consequently, the resulting vibrational Schr\"{o}dinger equation quickly becomes intractable already for moderately sized molecules in the presence of strong anharmonicity.
Furthermore, the Taylor expansion is an inherently local approximation, and therefore, it is inappropriate for double-well potentials or other more complex PESs for which there is no clear single reference geometry or for problems where the convergence of the Taylor series may be slow.

In the present work, we aim to overcome these drawbacks by utilizing a more general second quantization framework, which allows us to combine a generic $n$-mode PES expansion with the vDMRG algorithm in our novel $n$-mode vDMRG method.
The $n$-mode expansion, with its second-quantized form as proposed by Christiansen,\cite{christiansen2004_nMode-SecondQuantization} provides a framework that is better suited for strongly anharmonic systems, as it offers full flexibility in terms of both the form of the vibrational Hamiltonian and the choice of single-particle basis functions.
To demonstrate the generality of our $n$-mode vDMRG algorithm, we do not rely on any specific PES parametrization and instead compute the PES on the fly based on \textit{ab initio} electronic structure calculations during the basis set optimization.
We combine our on-the-fly PES construction with the vibrational self-consistent field (VSCF) algorithm\cite{Gerber1986_vSCF,Bowman1986,Hansen2010_vSCF} to obtain an enhanced, intrinsically anharmonic modal basis set for the $n$-mode vDMRG method.

In the following, we first present the theoretical basis of the $n$-mode vDMRG method in Section \ref{sec:theory}.
First, we briefly discuss the motivation for the $n$-mode potential energy surface expansion in Section \ref{subsec:pes} before presenting our VSCF algorithm utilized for calculating an optimized second quantization basis in Section \ref{subsec:vscf}.
We then present the $n$-mode vDMRG algorithm in Section \ref{subsec:vdmrg} and apply the sampling reconstruction of the complete active space (SRCAS) procedure for the \latin{a posteriori} analysis of the vibrational wave function in Section \ref{subsec:srcas}.
We provide the computational details of our $n$-mode DMRG-based vibrational framework in Section \ref{sec:comp_det} and then demonstrate its capabilities in Section \ref{sec:results} by applying it to methyloxirane, which is a suitable test case for both the PES construction and the subsequent vibrational structure calculation.

\section{Theory}
\label{sec:theory}

\subsection{Potential Energy Surface Construction}
\label{subsec:pes}

A general potential energy surface can be expressed in a high-dimensional model representation (HDMR) format.\cite{rabitz99_hdmr,alis01_hdmr,manzhos06_hdmr-nn-pes}
The basic idea behind the HDMR expansion method is to express a $(3N-6)$-dimensional PES as a sum of terms, each of which involves a subset of the $3N-6$ coordinates.
This approach results in a sum-over-terms form, in which terms are grouped together based on the number of degrees of freedom they depend on.
Hence, the potential assumes the form
\begin{equation}
\begin{aligned}   
\mathcal{V} = &\sum\limits_{i=1}^{3N-6} \mathcal{V}^{[i]}_1(r_i) \, + \sum\limits_{i<j}^{3N-6} \mathcal{V}^{[ij]}_2(r_i,r_j) \, + \sum\limits_{i<j<k}^{3N-6} \mathcal{V}_3^{[ijk]}(r_i,r_j,r_k)\\
&+ \ldots + \sum\limits_{i<j<k<\ldots < m}^{3N-6} \mathcal{V}_n^{[ijk\ldots m]}(r_i,r_j,r_k,\ldots ,r_m) \, ,
\label{eq:pes_hdmr}
\end{aligned}
\end{equation}
where $i,j,k, \ldots ,m$ each label one of the $3N-6$ degrees of freedom, and the $\mathcal{V}_n$ are functions dependent upon $n$ of the $3N-6$ coordinates $r_i,r_j,r_k, \ldots ,r_m$.
The $\mathcal{V}_n$ do not need to be in product form or, in fact, adhere to any specific functional format in general.
The HDMR expansion is based on the underlying assumption that the $(3N-6)$-dimensional potential can be accurately represented by a multitude of functions $\mathcal{V}_n$, each of which with a dimensionality $n$ lower than that of the entire system.
The power of this expansion is that, with an appropriate choice of functions $\mathcal{V}_n$ and coordinates, the series will converge quickly for many systems.\cite{rabitz99_hdmr,alis01_hdmr,manzhos06_hdmr-nn-pes}\\

The HDMR expression of the PES is commonly employed in vibrational calculations with the degrees of freedom corresponding to normal coordinates.
This yields the so-called $n$-mode expansion,\cite{Carter1997_VSCF-CO-Adsorbed,Bowman2003_Multimode-Code,Kongsted2006_nMode,Vendrell2007_ProtonatedWater-15D}, in which a PES can be written as
\begin{equation}
\mathcal{V} = \sum\limits_{i=1}^M \mathcal{V}_1^{[i]}(Q_i) + \sum\limits_{i<j}^M \mathcal{V}_2^{[ij]}(Q_i, Q_j) + \sum\limits_{i<j<k}^M \mathcal{V}_3^{[ijk]}(Q_i, Q_j, Q_k) + \ldots \, ,
\label{eq:pes_nmode}
\end{equation}
with terms $\mathcal{V}_n$ depending at most on $n$ of the $M$ normal modes at once.
The one-mode term $\mathcal{V}_1^{[i]}(Q_i)$ contains the variation of the potential upon changing the $i$-th normal coordinate $Q_i$.
Analogously, the two-mode terms $\mathcal{V}^{[ij]}(Q_i, Q_j)$ represent the PES variation for the simultaneous displacement along two coordinates $Q_i$ and $Q_j$.
The two-mode coupling terms are defined by
\begin{equation}
\mathcal{V}_2^{[ij]}(Q_i, Q_j) = \tilde{\mathcal{V}} \left( Q_i, Q_j \right) - \tilde{\mathcal{V}}\left( Q_i \right) - \tilde{\mathcal{V}}\left( Q_j \right) \hspace{2pt} , \label{eq:two-mode-term}
\end{equation}
where $\tilde{\mathcal{V}}\left( Q_i \right)$ and $\tilde{\mathcal{V}}\left( Q_i, Q_j \right)$ correspond to the potential energies of the structures where all coordinates except $Q_i$ and $Q_i,Q_j$, respectively, correspond to the reference structure.
In contrast to the nonrelativistic electronic Hamiltonian, vibrational mode interactions are not limited to one- and two-mode terms, as higher-order coupling terms appear, which are defined analogously to Eq.~(\ref{eq:two-mode-term}) in the $n$-mode expansion (cf., Refs.~\citenum{Kongsted2006_nMode} and \citenum{Toffoli2007_Automatic-PES}).
The exact potential is obtained by including all terms up to $\mathcal{V}_{M}^{[ij\ldots]}(Q_i, Q_j, \ldots)$, where $M$ is the total number of modes.
In practice, the expansion is truncated at the $n$th order term, with the truncation resulting in a hierarchy of approximations to the fully coupled PES.
As low-order terms often account for most of the vibrational correlation, the $n$-mode expansion often converges quickly with respect to $n$.

In this work, we choose to expand the PES along Cartesian normal coordinates because they are a natural choice for vibrational structure calculations on molecules with a well-defined equilibrium structure.
We therefore employ the $n$-mode PES expansion in this work, but our vDMRG algorithm can, in principle, straightforwardly be combined with HDMR PES expressions other than the normal-mode-based $n$-mode one.
By inserting the $n$-mode PES expansion into the vibrational Hamiltonian, we obtain in (mass-weighted) Cartesian normal coordinates
\begin{equation}
\mathcal{H} = \sum\limits_{i=1}^M \mathcal{T} (Q_i) + \sum\limits_{i=1}^M \mathcal{V}_1^{[i]}(Q_i) + \ldots + \sum\limits_{i<j<\ldots}^M \mathcal{V}_n^{[ij\ldots]}(Q_i, Q_j, \ldots) \, ,
\end{equation}
where $\mathcal{T}$ is the kinetic energy operator and rotational coupling terms have been neglected.\\

In contrast to a Taylor expansion of the PES, the $n$-mode PES representation is completely general, as it does not make any assumptions about the functional form of the individual contributions to the potential, therefore offering a more flexible representation of anharmonicity.
Furthermore, the $n$-mode expansion also allows for a second-quantized form of the Hamiltonian that relies on generic basis sets, as opposed to the harmonic oscillator-based canonical quantization that results from the Taylor series approximation.
Additionally, a less apparent benefit of the $n$-mode expansion emerges when it is combined with full configuration interaction (CI)-type solvers such as DMRG:
In canonical-quantization-based vDMRG calculations, an unfortunate complication can arise due to the very nature of the DMRG optimization.
Power series yield a rather inaccurate PES representation far from the reference geometry, especially in the presence of strong coupling between modes.
This can result in PES approximations for which $\mathcal{V} \rightarrow \pm \infty$ for large displacements.
Any truncated Taylor-based PES expansion will be dominated by the polynomials of maximum degree in that expansion at large displacements.
Depending on the (un)even order and the expansion coefficients, these polynomials by construction diverge at certain displacements, hence making the appearance of  so-called \quotes{holes} in the PES more likely.
Such a deficiency of the PES can cause the variational optimization to plunge into these unphysical minima, consequently resulting in unphysical solutions to the vibrational Schr\"{o}dinger equation.
These holes might not affect vibrational algorithms that are based on truncated configuration interaction, as such methods only explore a very limited region of the Hilbert space.
Holes are, however, encountered more frequently in vDMRG-based calculations, as the DMRG algorithm explores the entire Hilbert space more thoroughly.
Therefore, with the $n$-mode expansion for an efficient and flexible representation of the PES, we can also exploit the freedom of choice in the functional form of the expansion terms in Eq.~(\ref{eq:pes_nmode}) to circumvent functions which by default diverge to negative values at large displacements.

To demonstrate the generality of our approach, we in fact do not assume any kind of functional form of the potential terms, but construct the anharmonic single-mode potentials $\mathcal{V}_1^{[i]}(Q_i)$ and the mode-coupling potentials $\mathcal{V}_n^{[ij\ldots]}(Q_i, Q_j, \ldots)$ directly from electronic structure calculations:
First, the molecular structure is optimized, such that the PES is expanded around a physically relevant structure.
Second, we perform a normal mode analysis at the optimized geometry.
Third, the grid points for which the PES must be evaluated on the fly are determined through the second quantization basis set of choice.
Specifically for this work, we construct a numerical grid with displacements along each normal mode up to the $m$th harmonic inversion point $Q_i^{\text{max}} = \sqrt{\frac{2 m_i + 1}{\nu_i}}$, where $\nu_i$ is the harmonic frequency of the $i$th vibrational mode.
Hence, the displacements are mode-specific, such that the relevant region of the PES is covered by subsequent vibrational structure calculations for both high-energy and low-energy modes.
For a given normal coordinate $Q_i$, the grid points $Q_{\mu_i}$ are distributed equidistantly between the boundaries, that is $Q_{\mu_i} \in \left\lbrace -Q_i^{\text{max}}, -Q_i^{\text{max}} + \Delta Q_i, \dots , Q_i^{\text{max}} - \Delta Q_i, Q_i^{\text{max}} \right\rbrace$, where the step size is set to $\Delta Q_i = \frac{2 Q_i^{\text{max}}}{N_{\text{P}}-1}$ for $N_{\text{P}}$ grid points.
In the fourth step, we perform an electronic structure single-point calculation at each of the required grid points with the electronic structure model and basis set of choice to obtain the discrete PES representation.
And finally, in order to calculate the higher-order many-mode contributions to the PES, we repeat steps three and four with a simultaneous displacement of several modes.
This on-the-fly construction of the PES allows us to bypass any kind of fitting procedure that is often required when utilizing a pre-parametrized PES in an analytical format.

\subsection{Vibrational Self Consistent Field with Discrete Variable Representation}
\label{subsec:vscf}

Since HDMR PESs support second quantization based on generic basis sets, the $n$-mode expansion allows us to employ a basis set that is optimized to yield compact CI wave functions.
We therefore exploit modals obtained from a vibrational self-consistent field (VSCF)\cite{bowman86,hansen10} calculation for a compact representation of anharmonic vibrational modes.
For this reason, the PES construction is directly linked to the VSCF calculation in our framework to construct an optimized modal basis.
In close analogy to the Fock operator in electronic structure theory, we construct the vibrational mean-field operator $\mathcal{F}^{\textbf{k}}_i$
for each mode $i$ as
\begin{equation}
\mathcal{F}^{\textbf{k}}_i = h_i + \sum_{i\neq j} \langle \phi_j^{k_j} \vert \mathcal{V}_2^{[ij]} \vert \phi_j^{k_j} \rangle + \ldots + \sum_{i \neq j \neq \dots \neq m} \langle \phi_j^{k_j} \cdots \phi_m^{k_m} \vert \mathcal{V}_n^{[ij\dots m]} \vert \phi_j^{k_j} \cdots \phi_m^{k_m} \rangle \, ,
\end{equation}
where the one-mode terms $h_i$ comprise the kinetic energy operator and the one-mode potential of mode $i$, and $\textbf{k}$ denotes the set of vibrational quantum numbers $\lbrace k_i, k_j, \ldots, k_m \rbrace$ of the wave function.
The mean-field equations $\mathcal{F}_i^{\textbf{k}} \phi_i^{k_i} = \epsilon_i^{ k_i } \phi_i^{k_i }$ are solved self-consistently until convergence is reached for all one-mode functions $\phi_i^{k_i}$, which are commonly referred to as VSCF modals.
In practice, we solve the VSCF equations as a matrix eigenvalue problem through projection onto a finite basis set in which the VSCF modals are constructed.
Thus, modals are represented as linear combinations of $N_{\text{P}_i}$ functions $\left\lbrace \chi_{\nu} (Q) \right\rbrace_{\nu=1, \dots, N_{\text{P}_i}}$, referred to as the primitive basis, with
\begin{equation}
\phi_i^{k_i}(Q_i) = \sum_{\nu=1}^{N_{\text{P}_i}} c_{\nu}^{k_i}  \chi_{\nu} (Q_i) \hspace{2pt} .
\end{equation}
The mean-field operator in this basis reads as $\mathcal{F}_{\mu\nu} = \langle \chi_{\mu} \vert \mathcal{F} \vert \chi_{\nu} \rangle$.
The VSCF equation can therefore be written in matrix representation as
\begin{equation}
\textbf{FSc}=\epsilon\textbf{Sc} \hspace{2pt} ,
\end{equation}
where $\textbf{S}$ corresponds to the overlap matrix in the primitive basis.
By projecting the mean-field equations onto a finite basis set of size $N_{\text{P}_i}$ for each mode $i$, a set of $N_{\text{P}_i}$ eigenfunctions $\phi_i^{k_i}$ is obtained at each VSCF iteration, where we label the modals by their respective vibrational quantum number $k_i$.
Consequently, one of these solutions must be chosen to update the wave function of mode $i$, which in turn determines the mean-field potential for all other modes.
This selection makes it possible to target directly excited states by selecting an occupation number vector $\textbf{k}$ to follow during the self-consistent field procedure.
Note also that convergence of this procedure is fast (within a dozen iterations for the example studied in this work) so that we did not need to consider convergence acceleration protocols here.

The state-specific VSCF wave function is then obtained from the product \latin{ansatz} of the modals
\begin{equation}
\psi^{\textbf{k}} (\textbf{Q}) = \phi_1^{k_1}(Q_1)\phi_2^{k_2}(Q_2) \cdots \phi_M^{k_M}(Q_M) \hspace{2pt} ,
\end{equation}
resulting in an anharmonic wave function that takes the mode couplings into account in the mean-field limit.

Once convergence is reached, the state-specific VSCF energy can be calculated as
\begin{equation}
\begin{aligned}
E^{\textbf{k}} =& \, \langle \psi^{\textbf{k}} \vert \mathcal{H} \vert \psi^{\textbf{k}} \rangle\\
=& \, \sum_{i=1}^M \epsilon_i - \left[ \frac{1}{2!}\sum_{i\neq j} \langle \phi_i^{k_i} \phi_j^{k_j} \vert \mathcal{V}_2^{[ij]} \vert \phi_i^{k_i} \phi_j^{k_j} \rangle + \ldots \right.\\
& \, + \left. \frac{1}{n!} \sum_{i \neq j \neq \dots \neq m} \langle \phi_i^{k_i} \phi_j^{k_j} \cdots \phi_m^{k_m} \vert \mathcal{V}_n^{[ij\dots m]} \vert \phi_i^{k_i} \phi_j^{k_j} \cdots \phi_m^{k_m} \rangle \right] \hspace{2pt} ,
\end{aligned}
\end{equation}
where the additional terms introduced by the mean-field operator must be subtracted to avoid double counting of the potential interaction.

As a primitive basis for the VSCF modals, we choose the discrete variable representation (DVR) on a uniform grid,\cite{Colbert1992_DVR} as this basis is particularly well suited for a VSCF calculation conjoined with an on-the-fly PES construction.
The corresponding Fourier basis functions are localized about discrete values which are spread over a given interval, forming a DVR grid where each function only has a non-zero value at the point at which it is localized.
As a consequence, any multiplicative operator is of diagonal form in the DVR basis.
The major advantage of the DVR basis set is that it greatly simplifies the evaluation of the Hamiltonian matrix elements.
The kinetic energy operator elements can be calculated analytically, and the potential matrix elements are merely the value of the potential at the given DVR point,\cite{light00} so that no numerical integration is needed for the construction of the Fock matrix.
The corresponding kinetic and potential energy expressions of this DVR can be found in the Supporting Information.

As neither integral evaluations nor an analytical form of the PES are required with a DVR primitive basis, the PES constructed on the fly can directly enter the VSCF algorithm.
To that end, we choose a so-called Fourier DVR basis with a set of $N_{\text{P}_i}$ Fourier functions obtained for a given interval $\left[-Q_i^{\text{max}}, Q_i^{\text{max}} \right]$ which are equidistantly spread over the DVR grid.\cite{Colbert1992_DVR}
This corresponds to the very same grid as utilized for the $n$-mode PES expansion as described in Section \ref{subsec:pes}.
This choice of DVR points allows the on-the-fly calculation of the PES directly by electronic structure calculations only for the molecular geometries that are actually needed for the VSCF calculation.
We note here that we denote by the term "on-the-fly calculation" that the PES is constructed directly during the VSCF procedure through interfaces with electronic structure programs and that no fitting of the obtained energies is required, whereas this terminology is sometimes also used to indicate that the PES is continuously updated according to certain criteria based on an evolving wave function.
The latter approach has not been applied in the present work, but we might explore such an automated update of the PES guided by the wave function evolution in future work.

Through the joint on-the-fly PES construction and VSCF procedure the mean-field modals $\phi_i^{k_i}$ and corresponding single-particle energies $\epsilon_i^{k_i}$ for the $n$-mode Hamiltonian are obtained.
Apart from providing mean-field anharmonic vibrational frequencies and wave functions, these quantities can then also be utilized to construct a second-quantization framework for subsequent multiconfigurational vibrational structure calculations, such as vibrational CI or vDMRG.

We note here that, while we employ a directly conjoined grid construction procedure for the PES evaluation and the VSCF calculation in the present work in order to demonstrate the applicability of our methodology to arbitrary PES in HDMR format (without any requirements on the functional form or the need for fitting the PES), this strict equivalence of PES and VSCF grid points could be relaxed in future applications.
For instance, in order to reduce the number of electronic structure single-point calculations that are required, which currently scale as $\mathcal{O}\left(\binom{M}{n} \cdot N_{\text{P}}^n \right)$, the higher-order coupling terms could be calculated in a more coarse-grained manner by interpolating the electronic structure energies to augment the VSCF DVR basis size while lowering the computational cost of the PES construction.

\subsection{vDMRG in $n$-Mode Second Quantization}
\label{subsec:vdmrg}

We briefly review the theoretical foundations of vDMRG to prepare the grounds for our $n$-mode vDMRG method combined with an optimized basis set of anharmonic VSCF modals.

\subsubsection{Vibrational Density Matrix Renormalization Group}

As a variational method, vDMRG encodes the vibrational full CI wave function $ \vert \Psi \rangle$ as a matrix product state (MPS),
\begin{eqnarray}
  \vert \Psi \rangle &=& \sum_{\boldsymbol{\sigma}} c_{\boldsymbol{\sigma}} \vert \boldsymbol{\sigma} \rangle
  = \sum\limits_{\boldsymbol{\sigma}} \textbf{M}^{\sigma_1} \textbf{M}^{\sigma_2} \cdots 
    \textbf{M}^{\sigma_L}  \vert \boldsymbol{\sigma} \rangle \\
  &=& \sum\limits_{\sigma_1 \ldots \sigma_L} \sum\limits_{a_1 ... a_{L-1 }} 
    M_ {1 a_1}^{\sigma_1} M_ {a_1 a_2}^{\sigma_2} \cdots M_ {a_{L-1} 1}^{\sigma_L} 
   \vert \sigma_1 \rangle  \otimes \vert \sigma_2 \rangle  \otimes \cdots  \otimes \vert \sigma_L \rangle \, ,
\label{eq:mps}
\end{eqnarray}
where the CI coefficient $c_{\boldsymbol{\sigma}}$ for a given occupation number vector $\vert \boldsymbol{\sigma} \rangle = \vert \sigma_1 \cdots \sigma_L \rangle$ is obtained as the product of a set of matrices $\textbf{M}^{\sigma_l}$, one for each single-mode basis function $\sigma_l$ of the system.
In Eq.~(\ref{eq:mps}), the $M_{a_{l-1} a_{l}}^{\sigma_l}$ are rank-three tensors (except for the first and last ones in this tensor train) with the index $\sigma_l$ labeling the local state of the single-mode basis state at lattice site $l$.
The auxiliary indices $a_{l-1}$ and $a_{l}$ refer to the entries of the matrix $\textbf{M}^{\sigma_l}$, whose dimension will be truncated at a value $m$ that is known as the bond dimension.
The efficiency and accuracy of vDMRG depend crucially on the choice of the value for this bond dimension.

To match the MPS representation of the wave function, the Hamiltonian operator $\mathcal{H}$ can be written in a local decomposition, which is of matrix product form:
\begin{eqnarray}
  \mathcal{H} &=& \sum\limits_{\boldsymbol{\sigma}, \boldsymbol{\sigma}'} h_{\boldsymbol{\sigma}, \boldsymbol{\sigma}'} \vert \boldsymbol{\sigma} \rangle \langle \boldsymbol{\sigma}' \vert
  \label{eq:GenericOperator} = \sum\limits_{\boldsymbol{\sigma}, \boldsymbol{\sigma}'} \sum\limits_{b_1 \dots b_{L-1}}
                      H_{1 b_1}^{\sigma_1 \sigma_1'} H_{b_1 b_2}^{\sigma_2 \sigma_2'} \cdots H_{b_{L-1} 1}^{\sigma_L \sigma_L'} \vert
                      \sigma_1 \cdots \sigma_L \rangle \langle \sigma_1' \cdots \sigma_L' \vert \, ,
  \label{eq:MPO_0}
\end{eqnarray}
Here, the $H_{b_{i-1},b_i}^{\sigma_i, \sigma_i'}$ are now rank-4 tensors and no approximation is involved in adopting this format for any operator.
By contraction,
\begin{equation}
  H_{b_{l-1} b_l}^{[l]} = \sum\limits_{\sigma_l, \sigma_l'} H_{b_{l-1} b_l}^{\sigma_l \sigma_l'} \vert \sigma_l \rangle \langle \sigma_l' \vert \, ,
  \label{eq:MPOTensor}
\end{equation}
the Hamiltonian expression can be simplified to read 
\begin{eqnarray}
  \mathcal{H} &=& \sum\limits_{b_1 \dots b_{L-1}} H_{1 b_1}^{[1]} H_{b_1 b_2}^{[2]} \cdots H_{b_{L-1} 1}^{[L]} \\
&=& \textbf{\text{H}}^{[1]} \textbf{\text{H}}^{[2]} \cdots \textbf{\text{H}}^{[L]} \, ,
\label{eq:mpo}
\end{eqnarray}
a form that is known as a matrix product operator (MPO).
The matrices $\textbf{\text{H}}^{[l]}$ are operator-valued matrices that collect the elementary operators acting on a single-mode basis state at site $l$.

The expectation value $\langle \Psi \vert \mathcal{H} \vert \Psi \rangle$ of the Hamiltonian $\mathcal{H}$ (for normalized states $\Psi$) expressed as in Eq.~(\ref{eq:mpo}) over an MPS obtained from Eq.~(\ref{eq:mps}), is a non-linear functional of the entries $M_{a_{l-1} a_{l}}^{\sigma_l}$, which renders the simultaneous optimization of the coefficients of all tensors unfeasible.
If the energy is instead minimized solely with respect to the tensor centered on a given site $l$ while keeping all the other tensors fixed, a standard eigenvalue problem is obtained.\cite{Schollwoeck2011_Review-DMRG,Keller2015_MPS-MPO-SQHamiltonian}
The expectation value of the Hamiltonian $\langle \Psi \vert \mathcal{H} \vert \Psi \rangle$ can therefore be conveniently minimized in the MPS/MPO framework by sequentially optimizing the matrix coefficients of the individual single-particle basis functions in a sweeping procedure until convergence is reached.
The details regarding the implementation of the vDMRG algorithm can be found in Refs.~\citenum{Baiardi2017_VDMRG} and \citenum{Baiardi2019_HighEnergy-vDMRG}.

We showed in our previous work that converged vibrational energies can usually be obtained with a comparatively small maximum bond dimension with values of $m<100$ (a finding which is further confirmed by the results of the present paper).
In contrast to the wave function, the MPO representation of the Hamiltonian is not compressed as mentioned already above.
The bond dimensions $b_l$ of the MPO depend on the structure of the Hamiltonian.
As discussed, for example, in Refs.~\citenum{Frowis2010_MPOGeneric,Keller2015_MPS-MPO-SQHamiltonian,Hubig2017_MPOGeneric}, larger values of $b_i$ are required to encode longer-range interactions.
The MPO bond dimensions will therefore increase with the expansion order $n$ of the $n$-mode PES.
Hence, it is favorable that the HDMR PES expansion allows for efficiently encoding strong anharmonicity already with low-order many-body terms.

\subsubsection{$n$-Mode Second Quantization and $n$-Mode vDMRG}

For a vDMRG calculation, both the Hamiltonian and the wave function must be expressed in a second-quantized form obtained by projection onto a finite single-mode basis set.
For vDMRG with $n$-mode PES and generic modal bases, we leverage the $n$-mode second-quantization formalism introduced in Refs.~\citenum{christiansen2004_nMode-SecondQuantization} and \citenum{Wang2009_SQMCTDH}.
While our previously introduced Taylor-based vDMRG algorithm relies on the canonical quantization and therefore employs harmonic oscillator basis functions,\cite{Baiardi2017_VDMRG}
we consider here a general single-mode basis set that is different from the harmonic oscillator eigenfunction basis.
A basis for the full $M$-dimensional system can be constructed from all possible products of the single-mode functions $\phi_i^{k_i}$ as
\begin{equation}
  \psi_{k_1,\ldots, k_M} = \prod_{i=1}^M \phi_i^{k_i} \, .
  \label{eq:ManyBodyBasis}
\end{equation}

In the $n$-mode picture, we introduce a pair of creation and annihilation operators for each of the $N_i$ modal basis functions $\phi_i^{k_i}$ of each mode $i$.
The resulting $n$-mode occupation number vector (ONV) is given by
\begin{equation}
    | \textbf{n} \rangle = | n_1^1 , \, ... \, , n^{N_1}_1, \, ... \, , n^1_i, \, ... \, , n^{N_i}_i, \, ... \, , n^1_M , \, ... \, , n^{N_M}_M \rangle \, ,
    \label{eq:onv_nmode}
\end{equation}
where $n^{k_i}_i$ is the occupation of the $k_i$-th basis function $\phi_i^{k_i}$ associated with the $i$-th mode.
Any ONV describing a physically allowed state must fulfill the following three conditions:
\begin{equation}
  n^{k_i}_i \in \{0, 1\}, \hspace{.7cm} 
  \sum\limits_{k_i = 1}^{N_i} n^{k_i}_i = 1, \hspace{.7cm} \text{and} \hspace{.7cm}
  \sumover{i = 1}{L} \, \sum\limits_{k_i = 1}^{N_i} n^{k_i}_i = M \, . 
  \label{eq:onvcond}
\end{equation}
The first and second conditions imply that one and only one modal per mode can be occupied (as can be seen also from the fact that only one basis function per mode appears in Eq.~(\ref{eq:ManyBodyBasis})).
The third condition follows from the second one by summing over all possible modes and implies that the total occupation of the ONV is equal to the number of modes $M$.
Based on the $n$-mode-based ONV representation, the creation $\hat{a}_{k_i}^+$ and annihilation $\hat{a}_{k_i}$ operators can be introduced as\cite{christiansen2004_nMode-SecondQuantization}
\begin{eqnarray}
  \hat{a}_{k_i}^+ | \textbf{n} \rangle
    &=&  | \textbf{n} + \boldsymbol{1}_i^{k_i} \rangle  \, \delta_{0,n^{k_i}_i} \, ,
    \label{eq:CreationOperatorNMode} \\
  \hat{a}_{k_i} | \textbf{n} \rangle 
    &=&  | \textbf{n} - \boldsymbol{1}_i^{k_i} \rangle  \, \delta_{1,n^{k_i}_i} \, ,
    \label{eq:AnnihilationOperatorNMode}
\end{eqnarray}
where $\vert \boldsymbol{1}_i^{k_i} \rangle$ denotes the ONV with zero entries except for $n^{k_i}_i=1$.
As shown by Christiansen,\cite{christiansen2004_nMode-SecondQuantization} the Hamiltonian obtained from the $n$-mode PES given in Eq.~(\ref{eq:pes_nmode}) can be expressed in terms of the second-quantized operators defined above as follows:

\begin{equation}
 \begin{aligned}
  \mathcal{H} =  \sum_{i=1}^M \sum_{k_i,h_i=1}^{N_i} H_{k_i, h_i}^{[i]}
                 \hat{a}_{k_i}^+ \hat{a}_{h_i}
               + \sum_{i<j}^{M} \sum_{k_i,h_i=1}^{N_i} \; \sum_{k_j,h_j=1}^{N_j} H_{k_i k_j, h_i h_j}^{[i,j]}
			     \hat{a}_{k_i}^+ \hat{a}_{k_j}^+ \hat{a}_{h_i} \hat{a}_{h_j} \, ,
 \end{aligned}
 \label{eq:hamilton_nmode}
\end{equation}
where the one-mode integrals $H_{k_i, h_i}^{[i]}$ are calculated as
\begin{equation}
  H_{k_i, h_i}^{[i]} = \int\limits_{-\infty}^{+\infty} \phi_i^{k_i}(Q_i)
                       \left( \mathcal{T}(Q_i) + V_1^{[i]}(Q_i) \right) \phi_i^{h_i}(Q_i) \; \text{d} Q_i \, ,
  \label{eq:OneBodyIntegrals}
\end{equation}
and the two-mode integrals $H_{k_i k_j, h_i h_j}^{[i,j]}$ as
\begin{equation}
  H_{k_i k_j, h_i h_j}^{[i,j]} = \int\limits_{-\infty}^{+\infty}  \int\limits_{-\infty}^{+\infty} 
  \phi_i^{k_i}(Q_i) \phi_j^{k_j}(Q_j) V_2^{[i,j]}(Q_i, Q_j) \phi_i^{h_i}(Q_i) \phi_j^{h_j}(Q_j) \; \text{d} Q_i \text{d} Q_j \, .
  \label{eq:TwoBodyIntegrals}
\end{equation}
These expressions can be trivially extended to three-mode and higher-order terms based on the definition of the general, $n$-mode contribution of the potential given in Ref.~\citenum{Kongsted2006_nMode} and exploiting the theory presented in Ref.~\citenum{christiansen2004_nMode-SecondQuantization} to encode it in second quantization.

Note that the presence of coupling terms higher than two-body is a critical difference between electronic- and vibrational-structure calculations.
The Coulomb potential includes only pairwise interactions, and therefore, the electronic Hamiltonian contains only one- and two-body terms (\textit{i.e.}, strings with up to four different creation/annihilation operators at  most).
Vibrational Hamiltonians contain, in principle, up to $M$-body terms with $M$ being the number of modes of the molecule, and therefore, an appropriate choice of the reference coordinates and the single-mode basis set is crucial to compactly encode anharmonicity including mode couplings.\cite{Bulik2017_VSCFInternalOptimized,Jacob2009_LocalModes}
This is particularly relevant for vDMRG because, as we already mentioned, high-order coupling terms are difficult to encode in a compact MPO format.
In addition to the occurrence of higher-order terms, we also note that, in contrast with electronic structure theory, where the many-body bases given in Eq.~(\ref{eq:ManyBodyBasis}) must be properly antisymmetrized to take into account the permutational symmetry of the Hamiltonian, 
symmetrization is not needed for the $n$-mode vibrational Hamiltonian as the modes are distinguishable bosonic entities.
This is a consequence of the fact that the $n$-mode potential is clearly not invariant upon permutation of two different modes.
We will discuss the symmetry properties of the $n$-mode potential and the consequences thereof on the corresponding vDMRG algorithm in some detail below.

In the $n$-mode picture, every DMRG lattice site corresponds to one vibrational modal $\phi_i^{k_i}$.
The possible occupations of each modal are therefore one, if the modal $\phi_i^{k_i}$ is included in the many-body wave function, and zero otherwise.
Therefore, the local basis of each site on the DMRG lattice is two-dimensional.
As each modal is mapped to a site, the $n$-mode lattice has length $L=\sum\limits_i^M N_i$, where $N_i$ is the number of modals of the vibrational mode $i$, and $M$ is the number of modes.

\begin{figure}[htb!]
\includegraphics[width=0.9\textwidth]{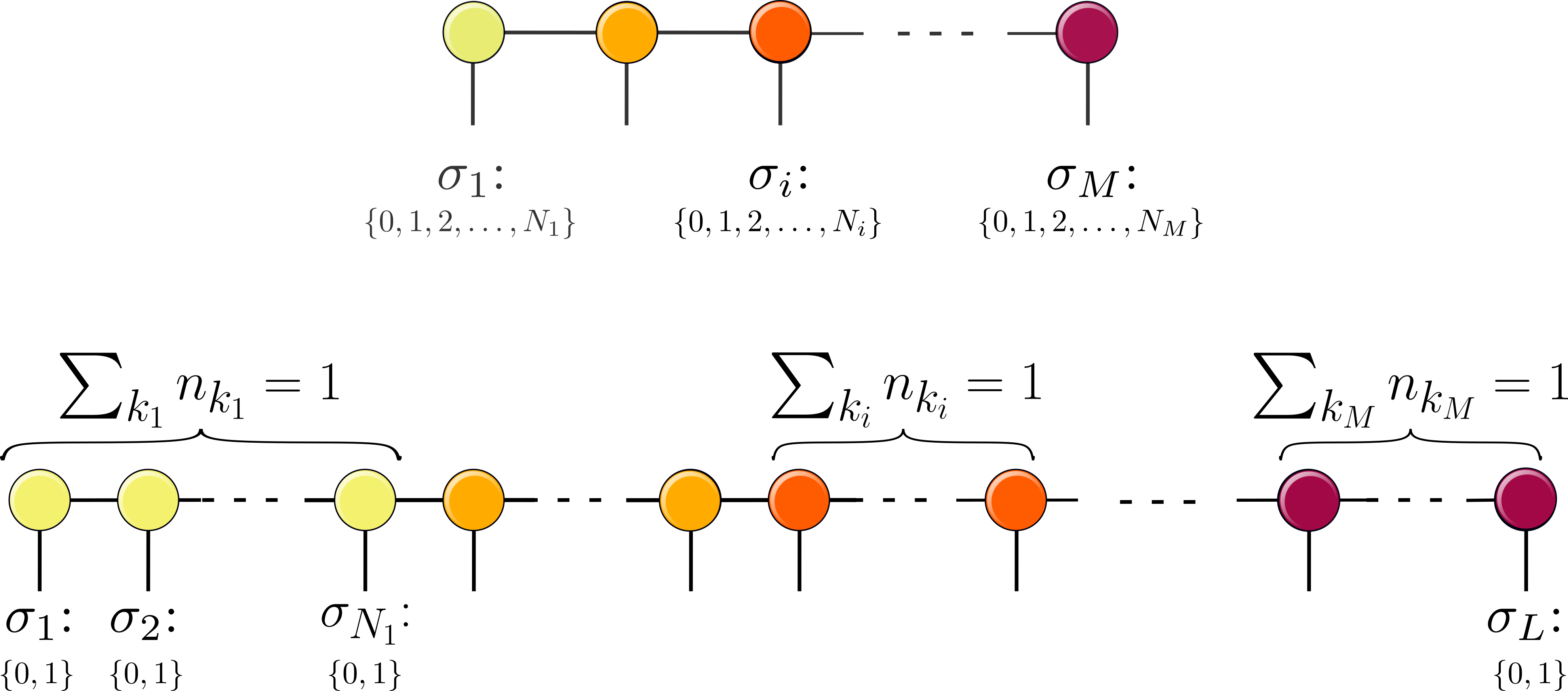}
  \caption{Vibrational MPS in canonical quantization (top) and $n$-mode quantization (bottom).
  Each circle of a given color represents an MPS tensor of a given vibrational mode.
  In our $n$-mode vDMRG algorithm, the modals of any given mode are grouped together, because this facilitates imposing the symmetry constraints which arise from the $n$-mode quantization, namely, that exactly one modal of each mode must be occupied in any ONV.}
  \label{fig:viblattice}
\end{figure}

We now briefly compare the original, harmonic oscillator-based, and the $n$-mode vDMRG lattice.
The two vibrational lattices, represented in the canonical quantization and in the $n$-mode picture are graphically compared in Fig.~\ref{fig:viblattice}.
In the canonical quantization picture, each vibrational mode is mapped to a single site on the vDMRG lattice, whereas the $n$-mode formulation maps each single-mode basis function to a lattice site.
As each mode is described by multiple modal basis functions, the $n$-mode lattice is significantly larger than the canonical vDMRG lattice.
This indicates that more sites need to be optimized in the sweeping procedure.
However, the local basis of size 2 in the $n$-mode picture is smaller than that in canonical vDMRG, where each site has $N_i$ possible states.
Furthermore, we recall that the relations given in Eq.~(\ref{eq:onvcond}) must hold. That is, one and only one modal per mode can be occupied in order for Eq.~(\ref{eq:onv_nmode}) to be physically acceptable.
This constraint can be expressed by stating that the number of particles for the $i$th mode, $n_i$, must be 1.
More formally, this implies that the number operator $\hat{n}_i=\sum\limits_{k_i=1}^{N_i}\hat{a}_{k_i}^+\hat{a}_{k_i}$ for any mode $i$ commutes with the Hamiltonian of Eq.~(\ref{eq:hamilton_nmode}).
The last property holds true since the Hamiltonian only contains strings of second-quantized operators with the same number of creators and annihilators per mode.
The Hamiltonian is therefore invariant under the action of the unitary group $\text{U}(1)$ for each mode and, therefore, under the action of the overall group $\text{NU}(1)$ that is defined as 
\begin{equation}
  \text{NU}(1) = \bigotimes_{i=1}^{N} \text{U}(1)^i \, .
  \label{eq:NU1_Def}
\end{equation}
As discussed, for example, in Refs.~\citenum{Vidal2008_MPS-Symmetric,Vidal2011_DMRG-U1Symm,Troyer2011_PEPS-Symmetry}, in the presence of quantum symmetries the gauge freedom in the definition of an MPS can be exploited to bring the tensors $M_{a_{i-1}a_i}^{\sigma_i}$ into a block-diagonal form.
The resulting structure of the MPS can be exploited to speed up the evaluation of its contraction with an MPO in the way as, in electronic-structure theory, the conservation of the $\alpha$ and $\beta$ orbitals can be imposed in the definition of an MPS.\cite{Keller2015_MPS-MPO-SQHamiltonian}
For $n$-mode vDMRG, the speed-up increases with the number of modes of the system.
This symmetry is a key difference of $n$-mode vDMRG compared to the original formulation of vDMRG, which does not possess any symmetries.

\subsubsection{VSCF-Based $n$-Mode vDMRG}

We already noted that the $n$-mode quantization is more suited to strongly anharmonic molecules than the original canonical quantization in vDMRG, as it allows both for a more flexible representation of the PES and for modal bases tailored to the degree of anharmonicity.
To leverage the latter, we now combine our $n$-mode vDMRG theory with the optimized VSCF modals which account for both the single-mode anharmonicity and the mean-field anharmonic mode couplings.
Therefore, the only remaining anharmonicity that has to be accounted for by the multiconfigurational nature of the MPS wave function is the correlation between the different modes.
This correlation can then be efficiently taken care of by the vDMRG algorithm.

\subsubsection{Excited-State Targeting with vDMRG[ORTHO] and vDMRG[FEAST]}

Since vDMRG variationally optimizes an MPS wave function, by default it  returns the multiconfigurational vibrational ground state and the corresponding anharmonic zero-point vibrational energy (ZPVE) $E_0$.
However, the quantities of interest in vibrational spectroscopy are not the ZPVEs, but the transition frequencies $\nu_k$ between different vibrational states, which can be calculated from the excited state energies $E_k$ as $h \nu_k = E_k - E_0$.
We therefore target the vibrationally excited states with excited state DMRG algorithms.
Various approaches have been developed to target excited state solutions with DMRG,\cite{Dorando2007_Targeted,Oseledts2016_VDMRG,Baiardi2019_HighEnergy-vDMRG} most of which can be straightforwardly applied to vDMRG.
In this work, we rely on two excited-state algorithms, namely vDMRG[ORTHO]\cite{Baiardi2017_VDMRG} and vDMRG[FEAST].\cite{Baiardi2022_DMRG-FEAST}

vDMRG[ORTHO] is a rather straightforward extension of the regular vDMRG algorithm, that optimizes excited states with a constrained optimization.
As all non-degenerate eigenstates of a Hermitian operator are mutually orthogonal, it is possible to target excited states by restricting the MPS optimization to the variational space orthogonal to all lower-energy states.\cite{McCulloch2007_FromMPStoDMRG,keller15,baiardi19b}
This requires a sequential calculation of the excited states, because the MPS of all lower-lying states must be available from a previous calculation.

While vDMRG[ORTHO] can be straightforwardly applied to low-lying excited states, for high-energy states not only does the computational effort increase, but the optimization may also become unstable. 
Therefore, we resort to vDMRG[FEAST] to target high-lying excited states.
vDMRG[FEAST]\cite{Baiardi2022_DMRG-FEAST} is based on the FEAST algorithm\cite{polizzi09}, which is an iterative subspace diagonalization algorithm for generalized eigenvalue problems.
The vDMRG[FEAST] algorithm simultaneously optimizes all states within a given target energy interval by leveraging the Cauchy integral theorem to numerically approximate the subspace projector.
Two key advantages of vDMRG[FEAST] make it particularly suited to calculate excited state MPS:
Firstly, only linear equations need to be solved in place of more involved eigenvalue problems.
Secondly, these linear systems are mutually independent and can therefore be solved in parallel.
Therefore, vDMRG[FEAST] can also be applied to energy regions with a high density of states, as the entire energy range can be calculated in a single step by trivial parallelization.
The vDMRG[FEAST] algorithm requires a lower and upper energy limit of the targeted energy interval as input, as well as the number of states to account for and an initial guess for each one of those states.

Ideally, the calculated excited states are fully converged with respect to both the number of optimization sweeps and the bond dimension of the wave function.
However, approximate transition frequencies can also be calculated between states that are not fully converged.
Such approximate states are orthogonal by construction if vDMRG[ORTHO] is employed as an excited-state targeting method, as the optimization is constrained to the parameter space orthogonal to all previously calculated states.
If vDMRG[FEAST] is applied, the orthogonality between the resulting states is ensured within a given subspace, as the final wave functions are obtained through diagonalization within the subspace.
The orthogonality between states in different energy intervals, or more generally between states obtained from different vDMRG[FEAST] calculations, can be monitored by evaluating the overlap between the two wave functions expressed as MPSs.

We note that, while we focused on the calculation of transition frequencies in the present work, the $n$-mode vDMRG method can also be applied to calculate other quantities of interest in vibrational spectroscopy.
For instance, the anharmonic dipole oscillator strengths can be evaluated within the $n$-mode vDMRG framework if a dipole surface is available.
A given property surface can be expressed in $n$-mode second quantization analogously to that of the PES. The resulting second quantized operator can be encoded as an MPO equivalently to that of the Hamiltonian, and it can therefore straightforwardly be applied to a vibrational wave function expressed as an MPS.

\subsubsection{MPS Initialization}
The choice of the initial MPS plays a key role in any vDMRG calculation, regardless of whether it is a regular ground state optimization or an excited state calculation.
While the DMRG algorithm may converge independently of the initial guess MPS, the convergence rate to the global minimum can be enhanced significantly by properly choosing the starting point for the optimization.
Hence, while the initial MPS can be constructed with constant or random coefficients, a physically more reasonable initial guess will result in faster and more robust convergence to the correct eigenstate.
Therefore, we initialize the MPS as the mean-field state corresponding to the target state.
For a ground state calculation, the MPS is thus initialized with the ONV where each mode occupies its lowest-energy VSCF modal, whereas excited states can be initialized in the corresponding target ONV of the mean-field reference state.
Since vDMRG[FEAST] requires multiple linearly independent guess MPSs, we initialize the targeted states as discussed above, while the remaining guesses are initialized in a random superposition of the overtones and combination bands with a mean-field energy included in the targeted energy interval.

\subsection{Vibrational Sampling Reconstruction of the Complete Active Space Algorithm}
\label{subsec:srcas}

To perform a spectroscopic assignment, vibrational states must be characterized as fundamentals, overtones, or combination bands according to their CI coefficients.
The CI coefficients are, however, not directly accessible from an MPS wave function for practical reasons: the curse of dimensionality does not allow one to construct all many-mode product basis states, which is the reason for resorting to DMRG in the first place.
For this reason, the characterization of a vibrational state expressed as an MPS requires a reconstruction of the most important many-mode product basis states of the full CI expansion.
The CI coefficients can be calculated by evaluating the overlap of the MPS with individual ONVs.
Then, a stochastic procedure for the reconstruction of the many-mode basis states of the CI expansion allows for efficient sampling of the relevant configurational space.
In this work, we therefore develop and implement a variant of the original sampling reconstruction of the complete active space (SRCAS) procedure\cite{boguslawski11_srcas}, here tailored to vibrational states.

In our algorithm, the configurational space is sampled with a Metropolis--Hastings Markov chain with the following steps:
\begin{enumerate}
\item A starting guess occupation number vector (ONV) is generated and its CI coefficient $C_{\text{curr}}$ is calculated by calculating its overlap with the MPS.
\item From the guess state, a randomly (de)excited state is generated. To thoroughly sample important regions of the CI space while ensuring that the entire Hilbert space remains accessible, the newly proposed ONV is drawn from a Poisson distribution centered on the current ONV.
\item The number of simultaneously (de)excited modes is controlled by imposing an acceptance criterion $\eta_{\text{accept}}$ on the proposed changes from step 2 based on a uniform random distribution in order to adjust the sampling speed across the Hilbert space.
\item Of the newly generated ONV, the CI coefficient $C_{\text{new}}$ is calculated, and the ONV is stored if $\left| C_{\text{new}} \right| > \eta_{\text{store}}$.
\item The current reference ONV is updated with the newly generated one with a probability $P= \min \left[ 1, \frac{\left| C_{\text{new}} \right|^2}{\left| C_{\text{curr}} \right|^2} \right]$.
\item Steps 2 to 5 are repeated until the CI expansion is sufficiently reconstructed as measured by $\sum_i \left| C_i \right|^2 > \eta_{\text{complete}}$.
\end{enumerate}

The two random draws of steps 2 and 3 combined result in a new ONV that is connected to the previous one to ensure thorough sampling of regions of interest.
At the same time, multiple changes of the occupation are allowed at once to ensure that the Markov chain can reach all areas of the Hilbert space.
This vibrational SRCAS (vSRCAS) algorithm can be applied not only to MPSs obtained from $n$-mode vDMRG calculations but to any kind of vibrational MPS for which the ONVs can be expressed as integer vectors with elements within a finite range, such as the canonical vDMRG.

\section{Computational Details}
\label{sec:comp_det}

We applied our $n$-mode vDMRG framework to methyloxirane, as this is a challenging system for \latin{ab initio} anharmonic calculations due to three main reasons:
(1) its size is challenging with 24 vibrational modes;
(2) the extent of anharmonic mode coupling is significant;
and (3) accidental resonances can occur.

On-the-fly PES construction and the corresponding VSCF calculation were implemented and performed in the \textsc{Colibri}\cite{colibri} program.
Currently, our framework supports $n$-mode expansions including up to third-order mode couplings.
Due to the high modularity of our software, an extension to higher orders is trivial but would yield a steep increase in the overall computational cost.
In this work, four different PESs were constructed by adopting a hierarchy of increasingly comprehensive $n$-mode expansions.
First, we calculated the 17-mode fingerprint region of methyloxirane with a two-mode PES, in which the anharmonic couplings including the lowest-energy mode, the methyl rotation, and the six highest-energy vibrations, namely, the C-H stretching vibrations, were neglected.
Therefore, only the diagonal anharmonicity was included for these modes.
This fingerprint region was then also targeted with a three-mode PES to investigate the impact of higher-order mode couplings.
Then a 23-dimensional two-mode PES was calculated, where only the methyl rotation was decoupled from the other modes, whereas vibrational couplings with the C-H stretching modes were included.
Finally, a fully coupled two-mode PES of all 24 vibrational modes of methyloxirane was constructed.

For the on-the-fly PES construction during the VSCF calculation, our framework was interfaced with various quantum chemistry programs via the \textsc{Scine/Utilities} module\cite{scine_utils} of our general
\textsc{Scine} framework that is open source and free of charge.
The methyloxirane PESs were constructed by restricted density functional theory 
(DFT) electronic structure calculations with the B3LYP exchange–correlation functional\cite{becke88_b3lyp_part1,lee88_b3lyp_part2} including Grimme’s D3 dispersion correction\cite{grimme10_d3dispcorr} and Becke--Johnson damping\cite{grimme11_d3bjdamping} with an aug-cc-pVDZ basis set\cite{kendall1992_augBasisSets} as implemented in the \textsc{Turbomole} program.\cite{turbomole}
While this DFT approach is not expected to yield accurate PESs, it serves our demonstration purposes and more reliable PESs can be obtained with more accurate \latin{ab initio} approaches if required.

The $5$th harmonic inversion point was chosen as the maximum displacement for all modes, and the number of DVR basis functions in the VSCF calculation was set to $N_P=11$ for each mode.
The adequate choice of these parameters has been determined through a numerical convergence analysis of a full-dimensional PES of water, which is tabulated in the Supporting Information.
The on-the-fly PES construction and VSCF calculation were parallelized with shared memory (OpenMP) parallelization for single-node calculations and with distributed memory (MPI) parallelization for multi-node infrastructure.
The latter option was leveraged for the calculation of the PES by activating up to 512 cores to parallelize the single-point calculations fully.

In addition to on-the-fly PES construction, \textsc{Colibri} can perform VSCF calculations based on precalculated grid point values or a sum-of-terms PES supplied as an input text file.
As the four PESs shared a large number of grid points, already calculated grid points were stored and reused, whereas new single-point calculations were performed on the fly when required in the VSCF calculation.
The DVR coefficients of the modals were initialized with the eigenvectors of the anharmonic one-mode Hamiltonian to accelerate the VSCF convergence.
A target state to be followed during the self-consistent field procedure must be specified.
To obtain the anharmonic mean-field transition energies, we carried out state-specific excited-state VSCF calculations.

The novel $n$-mode vDMRG algorithm has been implemented in our \textsc{QCMaquis} DMRG software package.\cite{qcmaquis}
\textsc{QCMaquis} supports two different optimization algorithms, namely, a single-site optimizer and a two-site optimizer, which optimize the tensor coefficients of either one or two sites of the lattice simultaneously.
As expected, the two-site optimization was computationally more costly but generally less prone to convergence to local minima.
However, we found the single-site optimizer combined with a perturbation-based subspace expansion to be sufficient for the vibrational structure problem considered in this work.

The $6$ lowest-energy modals of each mode as obtained from a ground-state VSCF calculation were chosen to construct the local basis of the $n$-mode vDMRG lattice for all vDMRG calculations.
Choosing a common modal basis set in which to expand all vibrational wave functions facilitates comparing different multiconfigurational wave functions and allows for straightforward excited-state calculations with vDMRG[ORTHO] and vDMRG[FEAST].

For all PESs, both vDMRG[ORTHO] and vDMRG[FEAST] calculations were performed with a bond dimension of $m=50$ and $n_s=50$ sweeps, which are denoted as ORTHO-L and FEAST-L accordingly, as we observed this parameter combination to yield vibrational energies converged to the cm$^{-1}$ level.
Additionally, vDMRG[FEAST] was investigated with a smaller bond dimension of $m=10$ and for only $n_s=10$ sweeps, which we denote as FEAST-S, in order to assess whether such a setting that leads to more cost-efficient calculations would deteriorate the accuracy of the large vDMRG[FEAST] calculation.

As our $n$-mode vDMRG algorithm is conceptually a variant of a generic vibrational configuration interaction (VCI) solver, we also implemented a simple VCI algorithm that relied on the same basis of anharmonic mean-field VSCF modals to allow for a direct comparison with the $n$-mode vDMRG calculations.
Our implementation of VCI in \textsc{Colibri} provides several options:\cite{colibri}
A specific reference ONV can be chosen around which the VCI space is expanded, where by default we take the mean-field reference ONV of the targeted state as a reference state, meaning that for excited-state calculations we expand around the corresponding excited ONV.
The maximum number of simultaneously (de)excited modes, as well as the maximum excitation degree of every individual mode and the maximum total (de)excitation degree, can also be set on input.
By default, our VCI calculations included all ONVs with one or two simultaneously (de)excited modes compared to the reference state.
(De)excitations with up to $10$ vibrational quanta in total with respect to the reference state were considered by default, where each mode could be excited up to the sixth-lowest energy VSCF modal, in agreement with our choice for the vDMRG lattice.
The corresponding VCI space is denoted as VCI(2,10) in the following, referring to the number of modes and vibrational quanta spanning the (de)excitation space.
Whenever applicable, we diagonalized the VCI matrix with a full divide-and-conquer eigensolver, whereas for large-scale VCI calculations we switched to the Davidson algorithm.
While VCI states obtained from a single VCI calculation are orthogonal by virtue of the diagonalization procedure, the orthogonality between states is no longer ensured when comparing states calculated within different VCI spaces.

\section{Results and Discussion}
\label{sec:results}

\begin{figure}[htb!]
\includegraphics[width=0.9\textwidth]{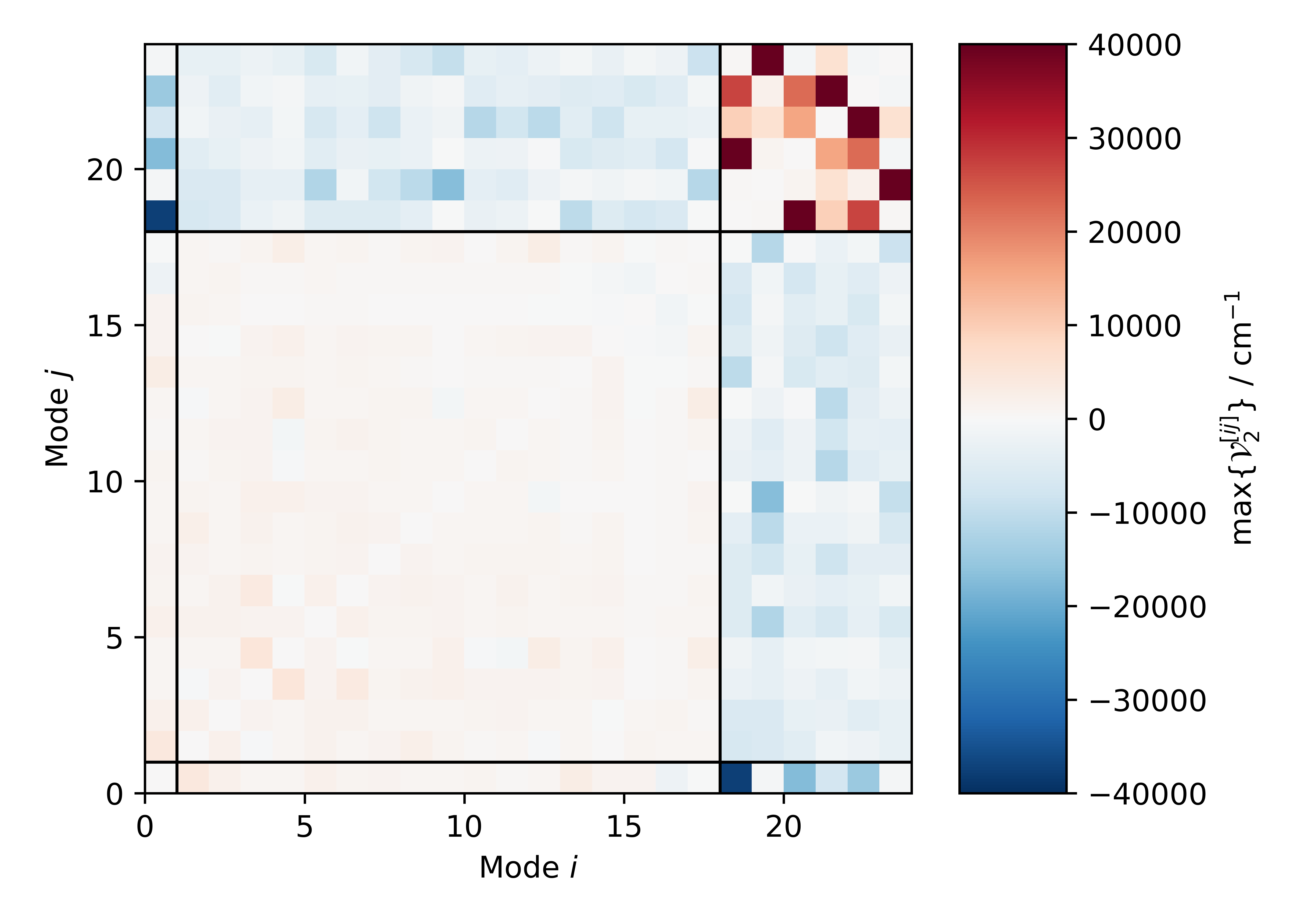}
\caption{Values of maximum magnitude of the two-mode coupling contributions for the on-the-fly constructed PES. Vibrational modes are ordered according to increasing harmonic frequencies from modes 1 to 24.}
\label{fig:twomodecouplinghist}
\end{figure}

\begin{figure}[htb!]
\includegraphics[trim={2.5cm 0.5cm 0cm 1.0cm}, clip, width=1.0\textwidth]{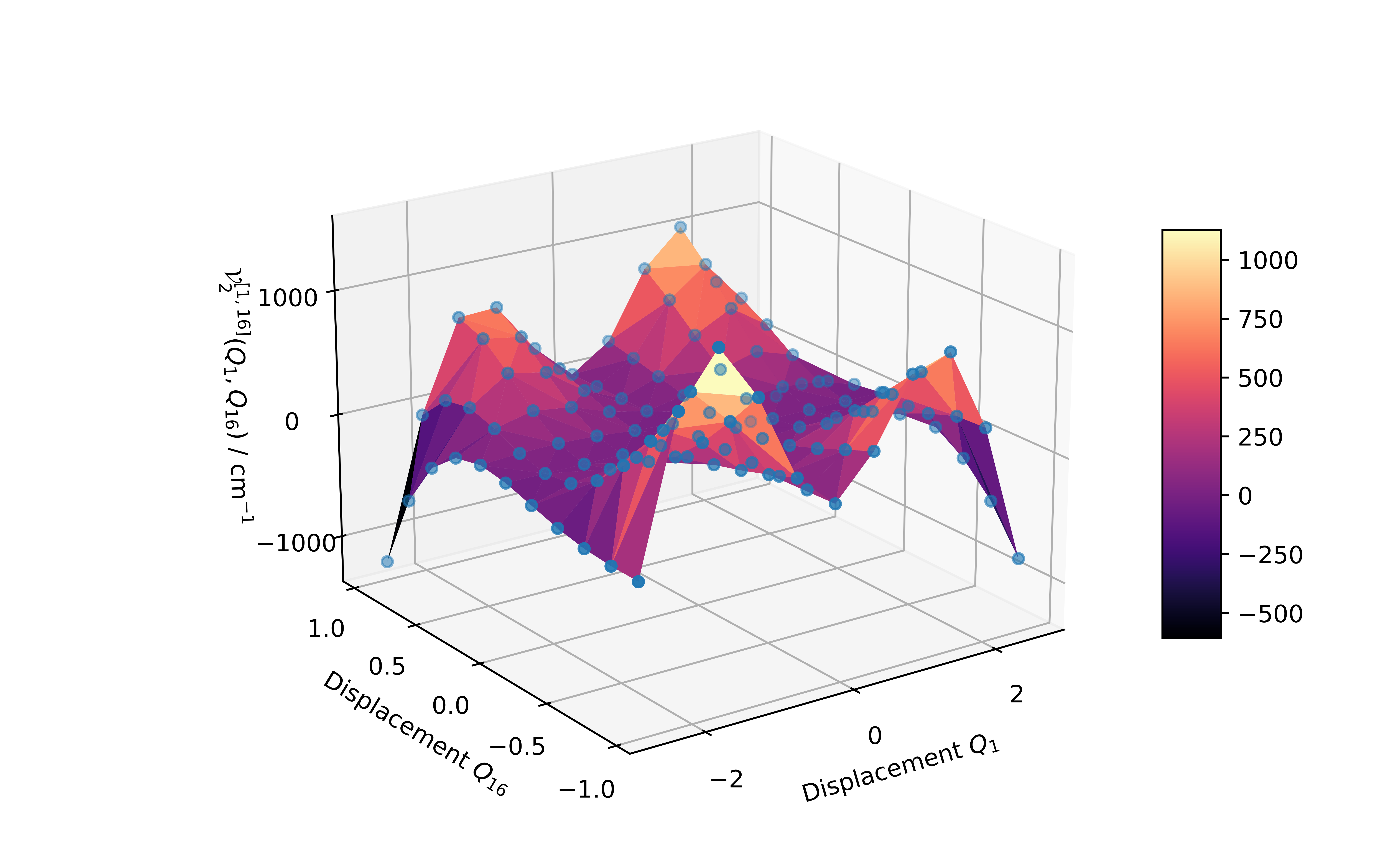}
\caption{Two-mode coupling potential between the methyl rotation (mode 1) and the hydrogen scissoring mode (mode 16) in methyloxirane with respect to the displacement along the dimensionless mass-weighted normal coordinates of the two modes. The blue points correspond to the on-the-fly calculated electronic structure single-points that directly enter into the VSCF algorithm.}
\label{fig:twomodecoupling}
\end{figure}

As shown in Figs.~\ref{fig:twomodecouplinghist} and \ref{fig:twomodecoupling}, the $n$-mode expansion of methyloxirane reveals a complex anharmonic potential energy surface.
The maximum two-mode coupling contributions contained within the displacements up to the $5$th harmonic inversion point span an energy range of roughly $100000$ cm$^{-1}$ and display characteristic coupling patterns.
In the fingerprint region, which in the following denotes modes 2 to 18, the maximum two-mode coupling contribution is comparatively small and predominantly positive.
The high-energy C-H stretching modes display a strong positive coupling with each other, while their maximum two-mode coupling with all other modes is negative.
The lowest-energy mode, which corresponds to the methyl internal rotation, is coupled strongly to the high-energy C-H stretching modes, which is to be expected, as several hydrogen atoms are either involved in both types of vibrations or in close proximity.
Such couplings between low-frequency large-amplitude modes and high-frequency X-H stretching vibrations are a direct consequence of the Cartesian-based description of the molecular vibrations.
The apparent block structure of the maximum two-mode coupling contributions justifies the hierarchical construction of the different PES expansions in this work, where starting from the fingerprint block, more coupling blocks can be included incrementally.

The anharmonic couplings not only span orders of magnitude in strength, but also display complex interaction patterns between different modes.
For instance, as expected due to the nature of the methyl rotation, its two-mode potential with the hydrogen scissoring features an intricate coupling landscape as shown in Fig.~\ref{fig:twomodecoupling}.
Since the PES energies evaluated at the grid points directly enter the evaluation of the $n$-mode terms in the VSCF algorithm, we bypass fitting the PES, and the anharmonicity displayed in both the one-mode potentials and the coupling terms between the modes can be accounted for exactly.

\begin{figure}[htb!]
\includegraphics[width=0.8\textwidth]{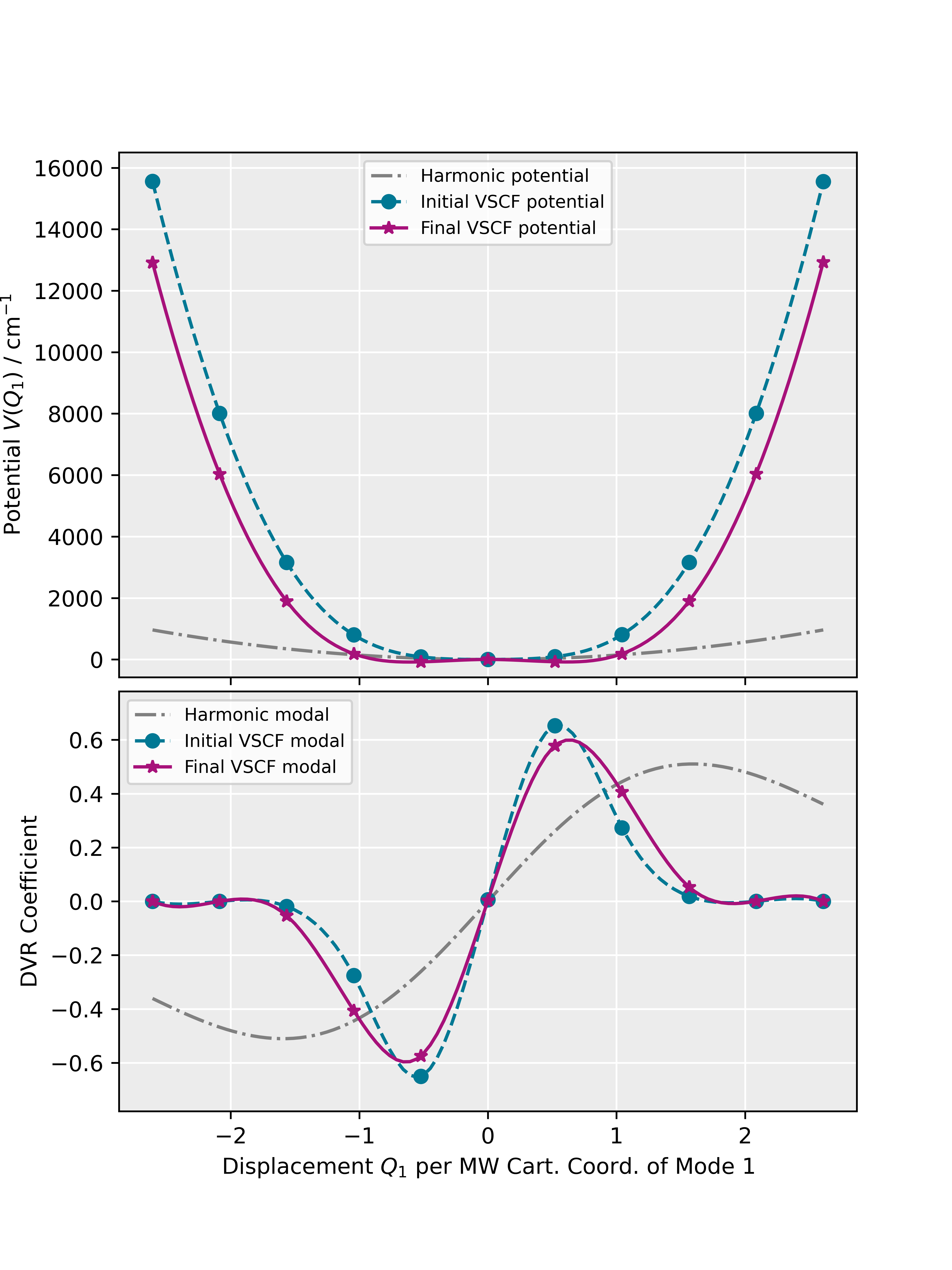}
\caption{Evolution of the VSCF potential (top) and corresponding modal (bottom) of the first fundamental excitation of methyloxirane during the VSCF calculation of the fully coupled 24-dimensional two-mode PES.}
\label{fig:vscfmode1}
\end{figure}

For each of the four on-the-fly calculated PESs, we obtained the VSCF ground-state wave function and energy.
Moreover, we calculated vibrational excitation energies with state-specific VSCF calculations.
As can be seen in Fig.~\ref{fig:vscfmode1}, the harmonic approximation of the lowest-energy mode fails to capture the sharp increase in the potential energy at large displacements, resulting in modals that are significantly too widespread for an accurate representation of the wave function.
During the VSCF calculation, the anharmonic potential including the mean-field coupling to all other modes is lowered as a result of the simultaneous optimization of the modals.
By initializing the DVR coefficients of the modals with the eigenvectors of the anharmonic one-mode problem, convergence was reached within a dozen VSCF iterations for all fundamental excitations described by the fully coupled 24-dimensional two-mode PES.

The ZPVE and the fundamental excitation frequencies were calculated within the harmonic approximation, with the DVR-VSCF procedure, with VCI, with the $n$-mode vDMRG algorithm, and with $n$-mode vDMRG[FEAST] for all four PESs.
The results for the fully coupled 24-dimensional two-mode PES are collected in Table~\ref{tbl:otf2}, whereas the three-mode fingerprint calculations are shown in Table~\ref{tbl:otf3_fp}.
The two-mode fingerprint results and the energies resulting from a 23-dimensional PES that neglects anharmonic couplings involving the methyl rotation can be found in the Supporting Information.

\begin{center}
\bgroup
\def\arraystretch{1.0}
\begin{table}[htb!]
	\centering
	\caption[2-full]{Comparison of harmonic, VSCF, VCI, and vDMRG energies for the fundamental excitations of methyloxirane ordered according to their harmonic energy.
    The potential energy surface considered is a two-mode on-the-fly calculated PES with all $24$ modes coupled.
    The C-H stretching vibrations marked by $\dagger$ cannot be unambiguously assigned, because there is no single dominating CI coefficient of those fundamentals in the calculated states.
    All energies are given in cm$^{-1}$ and the fundamental excitation energies are given relative to the corresponding ZPVE.}
    \label{tbl:otf2}
	\vspace{0.3cm}
	\begin{tabular}{lrrrrrr} 
        \hline\hline
	    State & Harmonic & DVR-VSCF & VCI(2,10) & FEAST-S & FEAST-L & ORTHO-L\\
		\hline
		ZPVE & 18586.2 & 18474.6 & 18380.5 & 18380.4 & 18375.1 & 18375.2\\
		$\nu_1$ & 194.7 & 293.3 & 215.9 & 207.8 & 205.1 & 205.1\\
		$\nu_2$ & 366.4 & 379.9 & 369.9 & 379.2 & 377.0 & 377.0\\
		$\nu_3$ & 411.3 & 426.4 & 420.4 & 423.4 & 421.3 & 421.3\\
		$\nu_4$ & 760.5 & 755.0 & 749.9 & 751.4 & 748.5 & 748.5\\
		$\nu_5$ & 837.0 & 827.0 & 822.3 & 825.2 & 822.1 & 822.0\\
		$\nu_6$ & 894.2 & 904.2 & 898.3 & 898.6 & 896.3 & 896.3\\
		$\nu_7$ & 964.1 & 958.4 & 951.5 & 956.8 & 954.7 & 954.7\\
		$\nu_8$ & 1026.0 & 1024.2 & 1019.9 & 1021.1 & 1019.6 & 1019.6\\
		$\nu_9$ & 1110.0 & 1105.4 & 1099.3 & 1103.2 & 1000.3 & 1100.3\\
		$\nu_{10}$ & 1139.2 & 1134.7 & 1128.4 & 1129.6 & 1128.7 & 1126.7\\
		$\nu_{11}$ & 1150.7 & 1141.0 & 1138.0 & 1138.8 & 1137.0 & 1137.1\\
		$\nu_{12}$ & 1175.3 & 1164.4 & 1161.6 & 1166.5 & 1164.3 & 1165.1\\
		$\nu_{13}$ & 1282.8 & 1267.9 & 1268.3 & 1266.0 & 1263.3 & 1263.4\\
		$\nu_{14}$ & 1381.5 & 1364.0 & 1357.0 & 1360.1 & 1357.3 & 1357.5\\
		$\nu_{15}$ & 1427.5 & 1407.5 & 1406.3 & 1409.0 & 1404.5 & 1404.6\\
		$\nu_{16}$ & 1454.3 & 1433.7 & 1433.6 & 1434.6 & 1433.4 & 1433.5\\
		$\nu_{17}$ & 1471.2 & 1449.7 & 1448.9 & 1450.1 & 1448.7 & 1448.7\\
		$\nu_{18}$ & 1516.8 & 1490.3 & 1484.7 & 1485.1 & 1477.0 & 1477.6\\
		$\nu_{19}$ & 3028.6 & 2906.4 & $\dagger$ & $\dagger$ & $\dagger$ & -\\
		$\nu_{20}$ & 3085.5 & 2951.1 & 2898.1 & 2900.6 & 2882.3 & -\\
		$\nu_{21}$ & 3096.7 & 2876.6 & 2818.7 & $\dagger$ & 2865.3 & -\\
		$\nu_{22}$ & 3099.8 & 2939.8 & 2960.3 & $\dagger$ & 2914.8 & -\\
		$\nu_{23}$ & 3120.3 & 2945.3 & 2830.5 & $\dagger$ & $\dagger$ & - \\		
		$\nu_{24}$ & 3177.8 & 2966.7 & 2932.7 & 2938.2 & 2931.7 & -\\
		\hline\hline
\end{tabular}
\end{table}
\egroup
\end{center}

Several trends can be observed when comparing ZPVEs obtained with the different vibrational structure methods and for the different potential energy surfaces.
With DMRG-based multiconfigurational methods, the anharmonic ZPVE is lower compared to the VSCF mean-field reference.
This variational decrease in energy becomes more apparent if the PES includes more coupled modes, as more anharmonic mode couplings are taken into account.
Similarly, the fundamental vibrational frequencies also decrease, with the magnitude of the reduction depending on the strength of the anharmonic many-mode couplings.
Whereas for the rather weakly coupled fingerprint region the anharmonicity, which is not captured by VSCF, only amounts to a few cm$^{-1}$, the frequencies of the lowest-energy mode and the high-energy modes differ substantially depending on the anharmonic vibrational structure method.
For modes that only exhibit weakly anharmonic features, a VCI calculation in a Hilbert space containing up to two simultaneously excited modes and up to ten-fold (de)excitations in total captures most of the multiconfigurational character.

However, for strongly coupled modes, vDMRG allows for a more thorough exploration of the relevant configuration space.
As can be seen by the slight decrease in ZPVE from VCI, a minimal MPS bond dimension of $m=10$ of the FEAST-S calculations already allows for comparable results by using the $n$-mode vDMRG algorithm.
The energies can be further improved by enlarging the bond dimension to $m=50$ while also increasing the number of sweeps for the DMRG optimization to ensure full convergence of the MPS coefficients, as was done for the FEAST-L and ORTHO-L calculations.

The proof-of-principle application of our novel vibrational framework to methyloxirane not only demonstrates the capabilities of the standard ground state $n$-mode vDMRG algorithm, but also showcases its extension to the DMRG[ORTHO] and DMRG[FEAST] excited-state methods.
Since we employed the very same mean-field basis set for all vDMRG[ORTHO] and vDMRG[FEAST] calculations, we can compare the methods for a given basis set.
Both excited-state algorithms can be applied to calculate the vibrational excitation energies in the fingerprint region, with the obtained frequencies being in excellent agreement with one another.
While FEAST-S often results in energies slightly higher than the corresponding ORTHO-L values, which indicates that the vibrational excitation energies are not fully converged with $m=10$, the FEAST-L frequencies are either identical to the ORTHO-L results to $0.1$ cm$^{-1}$ accuracy, or even marginally lower.
This small lowering of the FEAST energies when employing the very same bond dimension and number of DMRG sweeps for each MPS optimization as in ORTHO, is due to the fact that the subspace projection achieved through using Cauchy's integral theorem in the FEAST algorithm results in eigenstates that are obtained as a sum of optimized MPSs.
As the addition of two MPSs with bond dimension $m$ results in an MPS with a bond dimension of up to $2m$ if no compression is applied, the final FEAST MPS has a larger total bond dimension than the MPS optimized with conventional vDMRG.
Therefore, the vDRMG[FEAST] energy for a given choice of settings can be more accurate than the corresponding vDMRG[ORTHO] result.
However, as the DMRG algorithm converges rather quickly with respect to the bond dimension for many vibrational problems, both FEAST-L and ORHO-L can be converged up to the cm$^{-1}$ level of accuracy.

For low-lying excited states, the $n$-mode vDMRG[ORTHO] method is computationally more efficient since a single DMRG optimization has to be performed for each state.
However, this excited-state algorithm becomes inapplicable for high-lying states, because all lower-lying states are required as input for the orthogonality-constrained MPS optimization, including all possible overtones and combination bands.
Therefore, only vDMRG[FEAST] can target the high-lying C-H stretching vibrations (modes $19$ to $24$).

It should be noted, however, that vDMRG[FEAST] would allow for the use of different, state-specific mean-field basis sets for each targeted state, \latin{i.e.}, from excited-state VSCF calculations.
This tailoring of the basis set to the target state is not possible with vDMRG[ORTHO] because the same basis must be used for all sequential orthogonality-constrained calculations.
Additionally, due to the sequential nature of the ORTHO excited state calculation, the vDMRG[ORTHO] algorithm cannot be employed for a parallel calculation of multiple states.
With vDMRG[FEAST], however, several states can be targeted in parallel.

For the low-energy region of the vibrational spectrum, the FEAST energy interval could be set to rather several hundred cm$^{-1}$, whereas in the C-H stretching region we decreased the interval to $20$ cm$^{-1}$ because the density of states becomes very large.
While we only list the uniquely assignable fundamental excitation energies in Tables \ref{tbl:otf2} and \ref{tbl:otf3_fp}, we also obtained all overtones and combination bands within the targeted energy intervals because vDMRG yields eigenstates regardless of their spectroscopic character or accidental near-degeneracies.

\begin{figure}[htb!]
\includegraphics[width=1.0\textwidth]{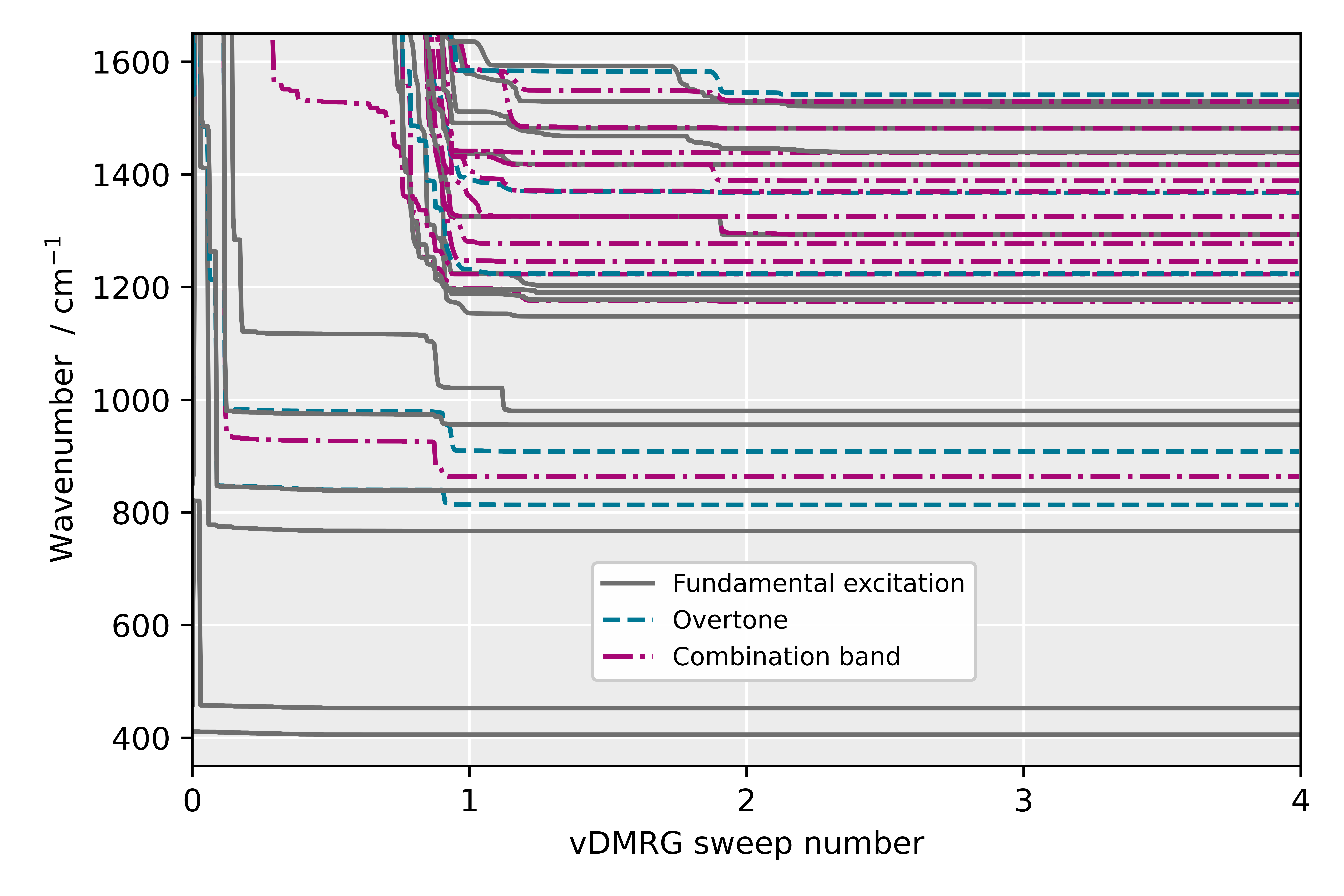}
\caption{$n$-mode vDMRG[ORTHO] optimization of the vibrational states in the fingerprint region of methyloxirane. The states are characterized by a vSRCAS for the optimized wave function and then classified as a fundamental excitation, overtone, or combination band based on their dominating CI coefficient.}
\label{fig:fp_conv_with_srcas}
\end{figure}

The obtained multiconfigurational vibrational states were assigned according to their dominating CI coefficient.
While the CI coefficients can be directly extracted from a VCI calculation, they were reconstructed with the vSRCAS algorithm in vDMRG.
As can be seen in Fig.~\ref{fig:fp_conv_with_srcas}, the vSRCAS procedure is vital for reliably assigning the eigenstates, because the density of states is high already in the fingerprint region, such that an assignment solely based on reference energies from harmonic, VSCF, or VCI calculations is no longer feasible.

While all fundamental excitations in the fingerprint regions could be trivially assigned with the CI coefficient reconstruction, this was not the case for the C-H stretching vibrations.
As can be seen from Fig.~\ref{fig:twomodecouplinghist}, these high-energy states are strongly coupled with each other, and the multiconfigurational wave functions obtained with VCI or $n$-mode VSCF cannot always be uniquely assigned to a specific reference ONV.
For instance, the multiconfigurational wave function obtained for the fundamental excitation of mode $19$ was not dominated by a single CI configuration.
This is to be expected for strongly coupled anharmonic systems, which yield such highly multiconfigurational states that cannot be calculated accurately with single-reference methods, but can instead be straightforwardly calculated with our $n$-mode vDMRG algorithm.
Generally, the bond dimension must be increased to accurately encode the strong couplings in the MPS wave function.
Even for the uniquely assignable C-H stretching modes, a very small bond dimension of $m=10$ no longer suffices to reach energy convergence, so that only for $m=50$ converged energies are obtained, as listed in Table \ref{tbl:otf2}.

\begin{figure}[htb!]
\includegraphics[width=1.0\textwidth]{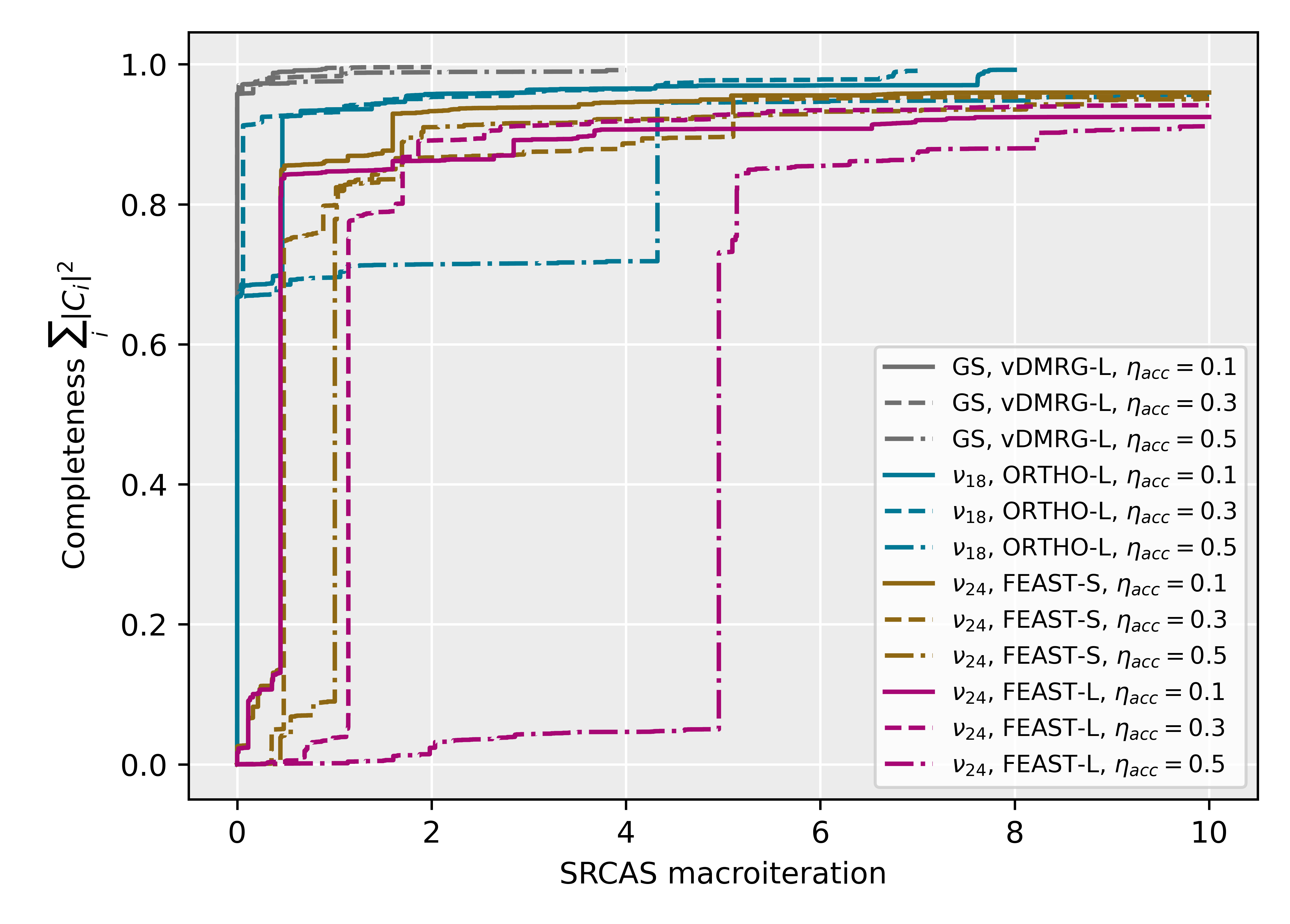}
\caption{vSRCAS convergence of the reconstruction completeness for different vibrational states and sampling settings.
The displayed ground-state wave function was obtained with regular $n$-mode vDMRG, the fundamental excitation of mode $18$ was calculated with vDMRG[ORTHO], and $\nu_{24}$ was targeted with vDMRG[FEAST].
The threshold for adding a sampled ONV to the CI reconstruction was set to $\eta_{\text{store}}=0.001$.
All vSRCAS procedures were stopped once a sampling completeness of $\eta_{\text{complete}}=0.99$ or 10 macroiterations of $10000$ (de)excitation trials each are reached.}
\label{fig:srcas_conv}
\end{figure}

In Fig.~\ref{fig:srcas_conv} the vSRCAS convergence is compared for different vibrational states obtained for the fully coupled 24-dimensional PES.
As expected, the number of sampling iterations needed to reach a certain target completeness of the CI coefficient reconstruction increased with the degree of multiconfigurational character of the wave function.
For the wave function at hand, the full Hilbert space contains over $10^{18}$ ONVs, such that a brute-force CI reconstruction is infeasible.
With our vSRCAS algorithm, the most important ONVs were sampled very efficiently, and the CI reconstruction accuracy could be directly monitored by the completeness measure.

The vSRCAS sampling efficiency across the Hilbert space can be easily adjusted by only accepting a fraction $\eta_{\text{accept}}$ of the proposed (de)excitations drawn from the Poisson distribution.
Acceptance ratios around $\eta_{\text{accept}}=0.3$ result in the most efficient overall sampling of the CI expansion for the system at hand.
The vSRCAS procedure converges to a reconstruction completeness of more than $90\%$ within 10 macroiterations for all 4 states displayed in Fig.~\ref{fig:srcas_conv}, regardless of the specific $\eta_{\text{accept}}$ sampling setting.
While for the vibrational ground state a completeness of more than $99\%$ could be reached within a few macroiterations, the excitations with stronger multiconfigurational character plateaued at a certain completeness.
This is due to the fact that no ONVs with CI coefficients below $\eta_{\text{store}}=0.001$ were added to the CI expansion.
By also accounting for contributions of ONVs with smaller weights, the required CI expansion completeness could be easily targeted by adjusting the corresponding vSRCAS settings.

\begin{center}
\bgroup
\def\arraystretch{1.0}
\begin{table}[htb!]
	\centering
	\caption{Comparison of harmonic, VSCF, VCI, and vDMRG energies for the fundamental excitations of methyloxirane ordered according to their harmonic energy.
    The PES considered was a 17-dimensional three-mode on-the-fly calculated PES which only treats the fingerprint region as coupled. Mode 1 and modes 19-24 were decoupled, as highlighted by the star $^\star$, and correspondingly the mean-field calculations already fully contained the one-mode anharmonicity, such that multiconfigurational calculations were omitted for these fundamentals.
    All energies are given in cm$^{-1}$ and the fundamental excitation energies are given relative to the corresponding ZPVE.}
    \label{tbl:otf3_fp}
	\vspace{0.3cm}
	\begin{tabular}{lrrrrrr} 
        \hline\hline
	    State & Harmonic & DVR-VSCF & VCI(2,10) & FEAST-S & FEAST-L & ORTHO-L\\
		\hline
		ZPVE & 18586.2 & 18802.0 & 18793.6 & 18792.7 & 18789.5 & 18789.5\\
		$\nu_1^\star$ & 194.7 & 381.4 & - & - & - & -\\
		$\nu_2$ & 366.4 & 402.3 & 395.8 & 395.3 & 394.9 & 394.9\\
		$\nu_3$ & 411.3 & 449.3 & 441.9 & 441.8 & 440.6 & 440.6\\
		$\nu_4$ & 760.5 & 767.5 & 760.2 & 757.0 & 755.6 & 755.1\\
		$\nu_5$ & 837.0 & 839.8 & 832.3 & 829.2 & 827.6 & 827.5\\
		$\nu_6$ & 894.2 & 953.3 & 939.8 & 933.3 & 933.0 & 931.4\\
		$\nu_7$ & 964.1 & 974.6 & 989.6 & 983.7 & 982.2 & 982.2\\
		$\nu_8$ & 1026.0 & 1065.4 & 1064.3 & 1062.9 & 1061.1 & 1061.1\\
		$\nu_9$ & 1110.0 & 1143.8 & 1144.4 & 1143.2 & 1141.6 & 1141.7\\
		$\nu_{10}$ & 1139.2 & 1187.6 & 1187.3 & 1183.3 & 1180.9 & 1181.0\\
		$\nu_{11}$ & 1150.7 & 1176.5 & 1171.1 & 1167.3 & 1164.8 & 1164.8\\
		$\nu_{12}$ & 1175.3 & 1195.2 & 1204.4 & 1200.2 & 1200.1 & 1200.2\\
		$\nu_{13}$ & 1282.8 & 1293.1 & 1288.8 & 1282.9 & 1283.2 & 1284.1\\
		$\nu_{14}$ & 1381.5 & 1407.9 & 1404.9 & 1403.4 & 1402.5 & 1402.5\\
		$\nu_{15}$ & 1427.5 & 1437.7 & 1438.1 & 1433.2 & 1432.5 & 1432.5\\
		$\nu_{16}$ & 1454.3 & 1463.8 & 1463.1 & 1463.9 & 1461.7 & 1461.7\\
		$\nu_{17}$ & 1471.2 & 1479.9 & 1478.9 & 1476.6 & 1478.9 & 1478.9\\
		$\nu_{18}$ & 1516.8 & 1528.4 & 1524.9 & 1521.2 & 1527.0 & 1527.1\\
		$\nu_{19}^\star$ & 3028.6 & 2990.7 & - & - & - & -\\
		$\nu_{20}^\star$ & 3085.5 & 3027.5 & - & - & - & -\\
		$\nu_{21}^\star$ & 3096.7 & 3164.7 & - & - & - & -\\
		$\nu_{22}^\star$ & 3099.8 & 3058.1 & - & - & - & -\\
		$\nu_{23}^\star$ & 3120.3 & 3166.2 & - & - & - & -\\		
		$\nu_{24}^\star$ & 3177.8 & 3246.5 & - & - & - & -\\
		\hline\hline
\end{tabular}
\end{table}
\egroup
\end{center}

\FloatBarrier

\section{Conclusions}
\label{sec:conclusions}

In this work, we introduced the $n$-mode vDMRG algorithm as a flexible method for calculating the vibrational spectrum of molecules even in the presence of strong anharmonicity.
The novel $n$-mode vDMRG algorithm allows for leveraging anharmonic modal basis sets to construct the DMRG lattice and can, therefore, be applied as an efficient full-CI solver for vibrational structure calculations on generic $n$-mode PESs.
We exploited the flexibility of the $n$-mode vDMRG method to work in a VSCF modal basis set whose optimization we combined with an on-the-fly $n$-mode PES construction.
We extended the ground-state $n$-mode vDMRG algorithm to two different excited-state algorithms and complemented the whole framework with an SRCAS CI coefficient reconstruction procedure tailored to vibrational wave functions for an efficient assignment of the eigenstates.
In the proof-of-principle application to the 24-dimensional PES of methyloxirane, we demonstrated the versatility of our novel framework.

We found the $n$-mode PES, which did not rely on a power-series expansion, in combination with anharmonic VSCF modals to be an adequate choice for molecular systems with a well-defined reference structure.
While normal coordinates are a natural choice for vibrational structure problems and are commonly used for HDMR PES expansions, any type of coordinates can be chosen to expand the PES in our vDMRG algorithm.
Due to our framework's modular and generic nature, the extension to different kinds of rectilinear coordinates could be realized straightforwardly.
This remains an open avenue for further exploration, and we might investigate alternative choices such as local mode coordinates\cite{Jacob2009_LocalModes,Cheng2014_LocalModesVCI,Panek2014_LocalModes,Rauhut2019_LocalModes} and precontracted basis functions\cite{Bowman1991_Precontraction,Carrington2002_Precontraction} in future work.

In contrast to the original vDMRG formulation, the novel $n$-mode vDMRG method provides complete flexibility with regard to the functional form of the vibrational Hamiltonian and the choice of single-mode basis functions.
We note here that the $n$-mode second quantization also features an additional flexibility over its harmonic-oscillator-based counterpart that is not exploited in the present work, which lies in the definition of the DMRG lattice.
In the original vDMRG formulation,\cite{Baiardi2017_VDMRG} where each lattice site corresponds to a vibrational mode, only the relative sorting of the modes can be changed.
Conversely, $n$-mode vDMRG, where each site of the lattice corresponds to a single basis function, allows for an arbitrary and more flexible sorting of the basis functions, therefore allowing for a more flexible choice of the lattice to further improve the DMRG convergence.
In future work, we will explore entanglement-based algorithms for determining the optimal modal selection and ordering on the vibrational lattice, as it is known from electronic structure that the lattice sorting can enhance the convergence rate of DMRG with respect to the bond dimension $m$.\cite{chan02,Legeza2003_OptimalOrdering,Legeza2003_Entanglement,Moritz2005_OptimalOrdering-DMRG,Legeza2016_OrbitalOptimization}
While the $n$-mode vDMRG algorithm yields, for the systems studied here, converged vibrational eigenstates already for low bond dimensions without any optimized lattice sorting (because a significant part of the anharmonicity can be directly encoded in the basis functions), an enhanced sorting on the vDMRG lattice will allow for an even more compact representation of anharmonic vibrational wave functions.
This compact representation of the vibrational wave function as a matrix product state could be further exploited by combining our $n$-mode framework with time-dependent DMRG algorithms,\cite{ren21_td-dmrg-nmode} where efficiently encoding the correlation within the wave function is of even greater importance to tame the so-called entanglement barrier effect.\cite{Legeza2019_OrbitalOptimization-TD,rams20_entanglementBarrier}
For this purpose, we plan to extend our $n$-mode framework to time-dependent vDMRG calculations in future work to also simulate complex anharmonic quantum dynamics.

\begin{acknowledgement}

The authors are grateful for financial support through the ‘Quantum for Life Center’ funded by the Novo Nordisk Foundation (grant NNF20OC0059939).

\end{acknowledgement}

%
%

\providecommand{\latin}[1]{#1}
\makeatletter
\providecommand{\doi}
  {\begingroup\let\do\@makeother\dospecials
  \catcode`\{=1 \catcode`\}=2 \doi@aux}
\providecommand{\doi@aux}[1]{\endgroup\texttt{#1}}
\makeatother
\providecommand*\mcitethebibliography{\thebibliography}
\csname @ifundefined\endcsname{endmcitethebibliography}
  {\let\endmcitethebibliography\endthebibliography}{}

\end{document}